\documentclass[twocolumn]{aastex63}
\usepackage{amssymb,amsmath}
\usepackage{graphicx}
\usepackage{xcolor,natbib}
\usepackage{multirow}
\usepackage[T1]{fontenc}
\usepackage{ae,aecompl}
\usepackage{newtxtext, newtxmath}
\usepackage{float}
\usepackage{subfigure}
\usepackage{longtable}
\usepackage{enumitem}
\usepackage{longtable,listings}
\usepackage[flushleft]{threeparttable}
\usepackage{parcolumns}
\usepackage{lineno}

\def\msig{$M_{\rm BH}- \sigma$\ }
\def\oiii{[O~{\sc iii}]\ }
\def\nii{[N~{\sc ii}]\ }
\def\sii{[S~{\sc ii}]$\lambda6716$\AA\ }

\def\oii{[O~{\sc ii}]$\lambda3727$\AA\ }

\submitjournal{ApJS}

\shorttitle{Larger $\sigma$ in Type-1 AGN?}
\shortauthors{Zhang XueGuang}

\begin{document}

\title{Are there larger stellar velocity dispersions in low redshift Type-1 AGN than 
in Type-2 AGN?}

\correspondingauthor{XueGuang Zhang}
\email{xgzhang@njnu.edu.cn}
\author{XueGuang Zhang$^{*}$}
\affiliation{School of Physics and technology, Nanjing Normal University, No. 1, 
Wenyuan Road, Nanjing, 210023, P. R. China}

\begin{abstract}
	The main objective is to check the Unified Model (UM) expected similar stellar velocity 
dispersions between Type-1 AGN and Type-2 AGN, then to provide further clues on BH mass properties. 
Not similar as previous comparisons of BH masses estimated by \msig relations to Type-2 AGN but 
Virial BH masses in Type-1 AGN, reliable stellar velocity dispersions $\sigma$ measured through 
absorption features around 4000\AA~ are directly compared between so-far the largest samples of 
6260 low redshift ($z~<~0.3$) Type-1 AGN and almost all the Type-2 AGN in SDSS DR12. Although half 
of Type-1 AGN do not have measured $\sigma$ due to unapparent absorption features overwhelmed by 
AGN activities, both properties of mean spectra of Type-1 AGN with and without measured $\sigma$ 
and positive dependence of $\sigma$ on [O~{\sc iii}] luminosity can lead to statistically larger 
$\sigma$ of all the Type-1 AGN than the 6260 Type-1 AGN with measured stellar velocity dispersions. 
Then, direct $\sigma$ comparisons can lead to statistically larger $\sigma$ in Type-1 AGN, with 
confidence level higher than 10sigma, after considering necessary effects of different redshift and 
different central AGN activities. Although Type-1 AGN have $\sigma$ only about $(9\pm3)$\% larger 
than Type-2 AGN, the difference cannot be well explained at current stage. Unless there was strong 
evidence to support different \msig relations or to support quite different evolution histories 
between Type-1 AGN and Type-2 AGN, the statistically larger $\sigma$ in Type-1 AGN provides a strong 
challenge to the Unified model of AGN.
\end{abstract}


\section{Introduction}

	The well-known constantly being revised Unified Model (UM) of AGN (Active Galactic Nuclei) 
has been widely accepted to explain most of different observational phenomena between broad line 
AGN (Type-1 AGN) and narrow line AGN (Type-2 AGN), due to effects of different orientation angles 
of central accretion disk \citet{an93}, combining with different central activities and different 
properties of inner dust torus etc. \citep{mb12, Oh15, ma16, bb18, bn19, kw21}. More recent reviews 
on the UM can be found in \citet{bm12} and in \citet{nh15}. Considering different viewing angles 
relative to central dust torus, the UM simply indicates that Type-2 AGN are intrinsically like Type-1 
AGN, but Type-2 AGN have their central accretion disk around black hole (BH) and broad line regions 
(BLRs) seriously obscured by central dust torus, leading to no optical broad line emission features 
in Type-2 AGN. The simple UM has been strongly supported by clearly detected polarized broad emission 
lines and/or clearly detected broad infrared emission lines for some Type-2 AGN \citep{mg90, hl97, 
tr03, nk04, or17, sg18, mb20}, and the strong resonance of silicate dust at 10${\rm \mu m}$ seen in 
absorption towards many Type-2 AGN but in emission in Type-1 AGN \citep{sh05}.

	However, even after necessary modifications to the UM, such as different properties of 
central dust torus and central activities, there are some other challenges to the UM. \citet{fb02} 
have supported different evolutionary patterns in Type-1 and Type-2 AGN. \citet{hi09} have shown higher 
average star formation rates in Type-2 AGN than in Type-1 AGN. More recently, \citet{vk14} have shown 
different neighbours around Type-1 AGN and Type-2 AGN. \citet{zy19} have shown that Type-1 AGN tend 
to have lower stellar masses of host galaxies than Type-2 AGN, through 2463 X-ray selected AGN in 
the COSMOS field. \citet{bg20} have discussed different host galaxy properties, such as UV/optical/IR 
colours and masses, together with differences in projected galaxy density at small scales (smaller 
than 100 kpc) and neighbouring galaxy properties, to favour an evolutionary scenario rather than a 
strict unified model in obscured and unobscured AGN. As detailed discussions in \citet{nh15}, the 
UM has been successfully applied to explain different features between Type-1 and Type-2 AGN in many 
different ways, however, there are many other features of structures/environments proved to be far 
from homogeneous among the AGN family.

	Based on commonly accepted framework of the UM, BH mass properties could be expected to be 
similar between Type-1 AGN and Type-2 AGN. As a pioneer work in \citet{nh09}, similar BH masses have 
been found between Type-1 AGN and Type-2 AGN with redshift from 0.1 to 0.2, although the main 
objective of \citet{nh09} is to check effects of radiation pressure force on gas dynamics in BLRs 
of AGN. BH masses in \citet{nh09} are estimated by the well-known \msig relations \citep{fm00, ge00} 
in Type-2 AGN but by the Virialization assumption \citep{ve02, pe04, sh11, rh11} applied in Type-1 
AGN. However, in recent years, \msig relations with much different slopes have been reported in 
different literature for different samples of objects,
\begin{equation}
\log(\frac{\rm M_{BH}}{\rm M_\odot})~=~\alpha~+~\beta~\times~
	\log(\frac{\sigma}{\rm 200~km\cdot~s^{-1}})
\end{equation}
The \msig relation with $\beta~\sim~4$ has been firstly reported in \citet{fm00, ge00}, based 
on dynamic measured BH masses and stellar velocity dispersions $\sigma$ of a small sample of 
nearby quiescent galaxies. More recent reviews of the \msig relations can be found in \citet{kh13}, 
\citet{mm13} and \citet{sg15} for samples of quiescent galaxies.

	Meanwhile, many studies have reported applications of \msig relations from quiescent 
galaxies to broad line AGN, such as results in \citet{bg05, ws13, hk14, wy15}. After well 
applications of reverberation mapping technique \citep{bm82} to determine BLRs sizes ($R_{BLRs}$, 
distance between BLRs and central BH) in the sample of reverberation mapped broad line AGN in 
AGNWATCH project (\url{http://www.astronomy.ohio-state.edu/~agnwatch}) \citep{pe04}, \citet{On04} 
have reported a scaling factor $f\sim5.5$ required to bring reverberation-based BH masses 
$M_{BH}=f~V^2R_{BLRs}/G$ into agreement with the quiescent galaxy \msig relationship. And then, 
\citet{wt10} have reported a virial factor as $f\sim5.2$ based on an updated reverberation sample 
including the low-mass local Seyfert 1 galaxies in the Lick AGN Monitoring Project (LAMP) 
(\url{https://www.physics.uci.edu/~barth/lamp.html}) \citep{bd10, ba15, wp18}. \citet{gr11} have 
reported a virial factor as $f\sim2.8$, based on an updated \msig relation of quiescent galaxies 
and an updated sample of AGN. \citet{pa12} have shown a preferred virial factor as $f\sim5.2$ based 
on a preferred forward statistical estimations. \citet{wy15} have reported a virial factor relative 
to full width at half maximum as broad line width, considering narrow-line Seyfert 1 galaxies. 
Moreover, considering the reverberation mapped broad line AGN, there are some improved \msig 
relations with much different slopes from 3.25 to 6.34 for different samples of objects listed 
and discussed in \citet{bt15, bb17, zh19, bt21}, which should lead to different BH masses of Type-2 
AGN with different \msig relations accepted.

	Based on the determined virial factors and the well-known R-L empirical relation for 
BLRs of reverberation mapped broad line AGN as discussed in \citet{kas00, bd13}, expression 
on virial BH masses of Type-1 AGN is being improved in common broad line AGN (not only in 
reverberation mapped broad line AGN), under the Virialization assumptions accepted to broad 
line emission clouds of central BLRs,
\begin{equation}
\log(\frac{\rm M_{BH}}{\rm M_\odot})=A+
	B\log(\frac{\rm \lambda L_{5100\AA}}{\rm 10^{44}erg\cdot~s^{-1}})+ 
	2\log(\frac{\rm V}{\rm 10^3km\cdot~s^{-1}})
\end{equation} 
where $V$ represents Kepler velocity of broad line emission clouds which can be traced by 
broad emission line width, and $B\log(\frac{\rm \lambda L_{5100\AA}}{\rm 10^{44}erg\cdot~s^{-1}})$ 
shows hints of distance of BLRs ($R_{\rm BLRs}$) to central BH. Since the empirical R-L 
relation has been firstly reported by \citet{kas00} with $B\sim0.7$ through variabilities of 
broad Balmer emission lines of the 17 nearby PG quasars, the R-L relation has been modified to 
$R_{\rm BLRs}~\propto~\lambda L^{\sim0.5}$ by more reverberation mapped broad line AGN with 
necessary corrections of host galaxy contaminations in \citet{bd13}. The well accepted 
$B~\sim~0.5$ different from the one used in \citet{nh09} should lead to different BH masses 
of Type-1 AGN. And moreover, there are different scale factors of $A$ in different literature. 
Meanwhile, for broad line AGN with measured stellar velocity dispersions, such as the AGN in 
\citet{bt15}, the virial BH masses have large scatters in \msig space, leading to large uncertainties 
on statistical comparisons of virial BH masses of Type-1 AGN under the Virialization assumption and 
the BH masses of Type-2 AGN by the \msig relations.

	Moreover, some independent methods have been proposed and applied to estimate central 
BH mass of individual AGN. \citet{bt11} have presented a direct BH mass measurement in the AGN 
Arp 151, based on motions of the gas responsible for the broad emission lines. \citet{pa14} have 
reported central BH mass measurements that does not depend upon the virial factor of five Seyfert 
1 galaxies from the LAMP 2008 sample, by directly modelling the AGN continuum light curves and the 
broad H$\beta$ line profiles. And more recently, \citet{wp18} have reported central BH mass 
measurements of seven Seyfert 1 galaxies from the LAMP 2011 sample, by re-constructing dynamic 
structures of central BLRs. However, the proposed independent method to estimate central BH masses 
of AGN can not be widely applied in normal AGN with single-epoch spectra.

	Different slopes in \msig relations and different factors in Virialization assumptions 
can lead to different properties of estimated BH masses in Type-2 AGN and in Type-1 AGN. It is 
necessary and interesting to re-check BH mass properties between large samples of Type-1 AGN and 
Type-2 AGN by different but direct methods, which is the main objective of the manuscript. In the 
manuscript, rather than the \msig relations applied to Type-2 AGN and the Virialization assumptions 
applied to Type-1 AGN, measured stellar velocity dispersions $\sigma$ are directly compared between 
large samples of Type-2 AGN and Type-1 AGN from SDSS DR12 (Sloan Digital Sky Survey, Data Release 
12, \citet{al15}). Section 2 presents data samples of Type-1 AGN and Type-2 AGN, methods to measure 
stellar velocity dispersions through absorption features around 4000\AA. Section 3 shows effects of 
AGN continuum emissions and broad line emissions on our measured stellar velocity dispersions in 
Type-1 AGN. Section 4 shows reliability of the measured stellar velocity dispersions. Section 5 
shows the main results and necessary discussions. Section 6 gives the final summaries and conclusions. 
And in the manuscript, the cosmological parameters of $H_{0}~=~70{\rm km\cdot s}^{-1}{\rm Mpc}^{-1}$, 
$\Omega_{\Lambda}~=~0.7$ and $\Omega_{m}~=~0.3$ have been adopted.

\section{Data Samples and Methods to Measure Stellar Velocity Dispersions}

\begin{figure*}  
\centering\includegraphics[width = 18cm,height=10cm]{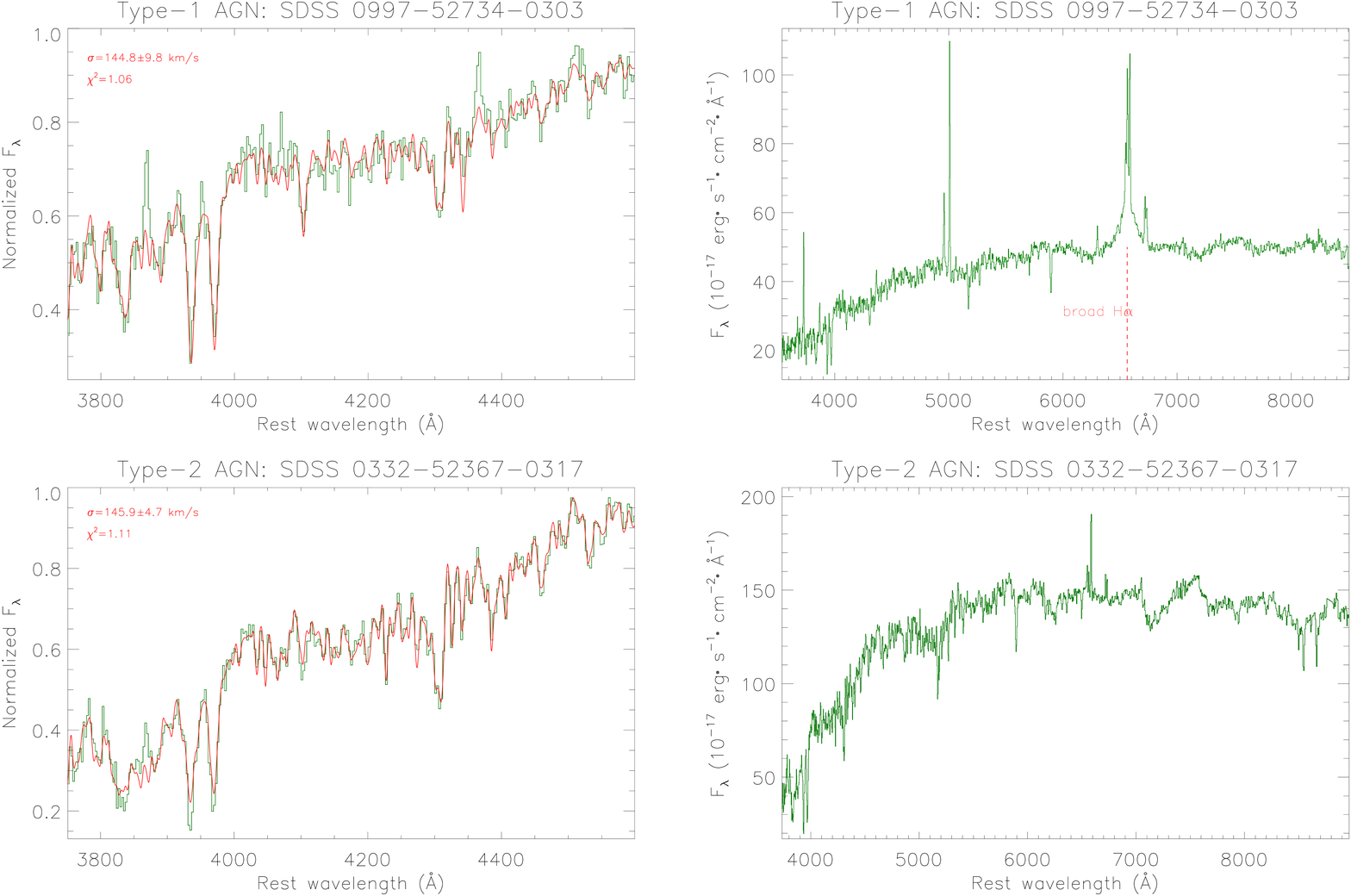}
\caption{Two examples on the best fitting results (solid red line) to the absorption features 
around 4000\AA~ (solid dark green line) in Type-1 AGN 0997-52734-0303 (PLATE-MJD-FIBERID) (top 
panels) and in Type-2 AGN 0332-52367-0317 (bottom panels). In each left panel, the measured 
stellar velocity dispersion $\sigma$ and corresponding calculated $\chi^2$ value (summed squared 
residuals divided by degree of freedom) are marked in top left corner. In right panels, solid 
dark green lines show the whole observed optical spectrum of the Type-1 AGN 0997-52734-0303 and 
the Type-2 AGN 0332-52367-0317. The apparent broad H$\alpha$ is marked in the top-right panel in 
the Type-1 AGN 0997-52734-0303.}
\label{sigma}
\end{figure*}

\begin{figure*} 
\centering\includegraphics[width = 18cm,height=11cm]{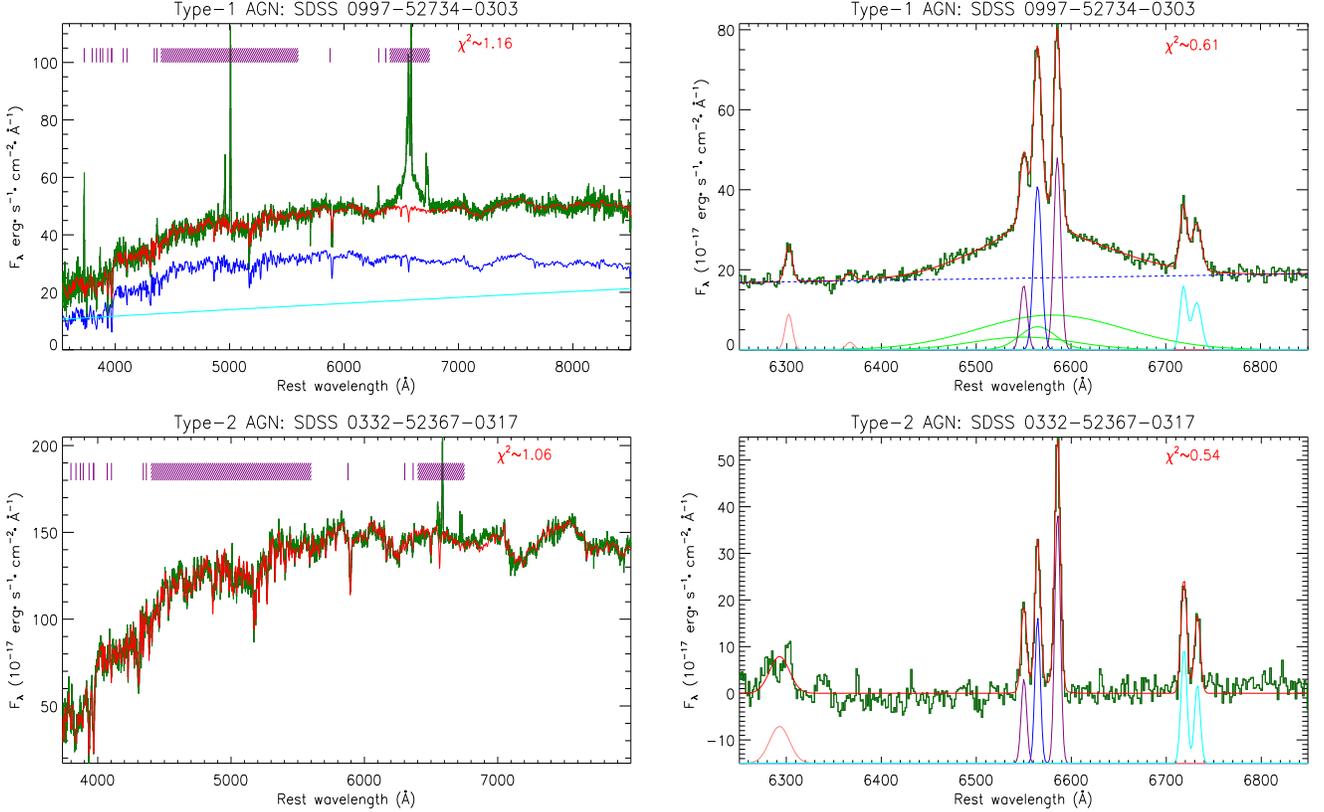}
\caption{Left panels show the SSP method determined best descriptions (solid red line) to the 
SDSS spectra (solid dark green line) of Type-1 AGN 0997-52734-0303 and Type-2 AGN 0332-52367-0317 
shown in Fig.~\ref{sigma}. In each left panel, from left to right, the vertical purple lines point 
out the emission lines being masked out when the SSP method is running, including \oii, H$\theta$, 
H$\eta$, [Ne~{\sc iii}]$\lambda3869$\AA, He~{\sc i}$\lambda3891$\AA, Ca~K, [Ne~{\sc iii}]$\lambda3968$\AA, 
Ca~H line, [S~{\sc ii}]$\lambda4070$\AA, H$\delta$, H$\gamma$, [O~{\sc iii}]$\lambda4364$\AA, 
He~{\sc i}$\lambda5877$\AA\ and [O~{\sc i}]$\lambda6300,6363$\AA\ doublet, and the area filled 
by purple lines around 5000\AA\ shows the region masked out including the optical Fe~{\sc ii} lines, 
broad and narrow H$\beta$ and \oiii doublet, and the area filled by purple lines around 6550\AA\ 
shows the region masked out including the broad and narrow H$\alpha$, \nii and \sii doublets. In 
top left panel, solid blue line shows the determined host galaxy contributions, solid cyan line 
shows the determine AGN continuum emissions in the Type-1 AGN. Right panels show the best descriptions 
(solid red line) to the emission lines around H$\alpha$ (solid dark green line) after subtractions 
of host galaxy contributions. In each right panel, solid blue line shows the determined narrow 
H$\alpha$, solid purple lines show the determine [N~{\sc ii}] doublet, solid pink lines show the 
determined [O~{\sc i}] doublet, solid cyan lines show the determined [S~{\sc ii}] doublet. In top 
right panel, solid green lines show the determined broad Gaussian components in the broad H$\alpha$, 
dashed blue line shows the determined power law continuum emissions underneath the emission lines. 
}
\label{line}
\end{figure*}

\begin{figure*}  
\centering\includegraphics[width = 18cm,height=6cm]{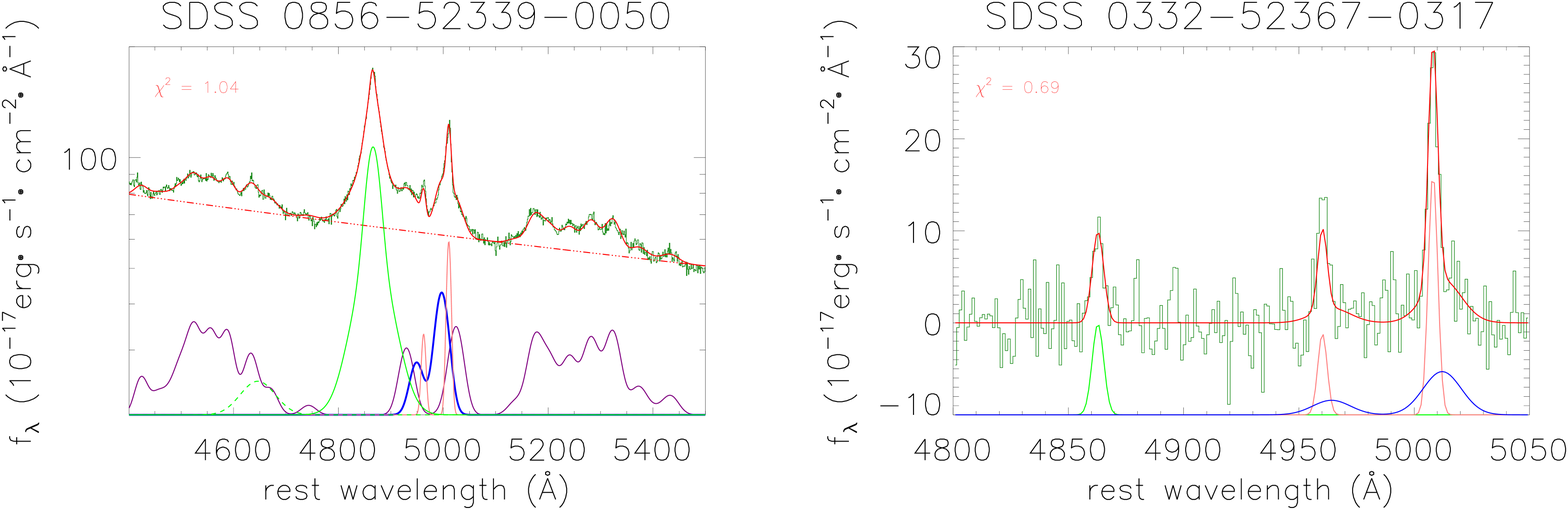}
\caption{Left panel shows the best fitting results (solid red line) to emission lines around H$\beta$ 
(solid dark green line) including apparent optical Fe~{\sc ii} emission features in the Type-1 AGN 
0856-52339-0050. Double-dot-dashed red line shows the determined power law continuum emissions, 
solid green line shows the determined broad H$\beta$, solid purple lines show the determined optical 
Fe~{\sc ii} lines, dashed green line shows the determined broad He~{\sc ii} line, solid pink lines 
show the determined core \oiii components, and thick blue solid lines show the determined broad 
blue-shifted \oiii components. Right panel shows the best fitting results (solid red line) to the 
emission lines around H$\beta$ (solid dark green line) in the Type-2 AGN 0332-52367-0317 of which 
SDSS spectrum is shown in Fig.~\ref{sigma} and Fig.~\ref{line}, after subtractions of host galaxy 
contributions. Solid green line shows the determined narrow H$\beta$, solid lines in pink and in 
blue show the determined core and extended components of [O~{\sc iii}] doublet. And the calculated 
$\chi^2$ values for the best fitting results are marked in the top-left corners in the panels.
}
\label{hb}
\end{figure*}

\begin{figure*} 
\centering\includegraphics[width = 18cm,height=6cm]{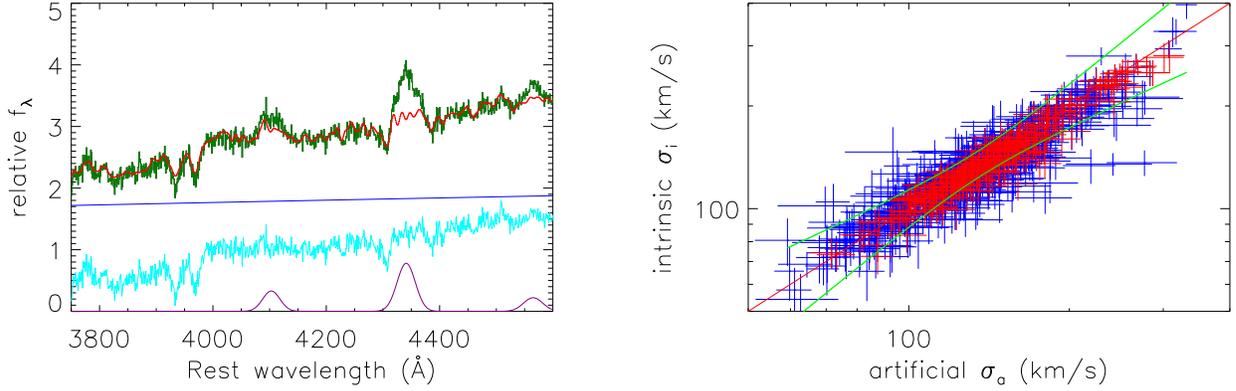}
\caption{Left panel shows an example of artificial spectrum around 4000\AA~ (solid line in dark 
green) considering contributions of AGN continuum emissions (solid blue line) and broad line 
emissions (solid purple lines), and the corresponding best fitting results (solid red line). 
In left panel, solid cyan line shows the spectrum $S_2(\lambda)$ around 4000\AA~ collected from 
the Type-2 AGN SDSS 0789-52342-0246. Right panel shows the correlation between intrinsic $\sigma_i$ 
and artificial $\sigma_a$. In right panel, symbols in blue are for all the data points with 
artificial $\sigma_a$ 5 times larger than their corresponding uncertainties, and symbols in red 
are for the high quality data points with artificial $\sigma_a$ 10 times larger than their 
corresponding uncertainties and $\chi^2$ smaller than 2, solid red line shows X=Y, solid green 
lines show the 99.99\% confidence bands to the linear correlation X=Y.
}
\label{sax}
\end{figure*}

\begin{figure} 
\centering\includegraphics[width = 8cm,height=15cm]{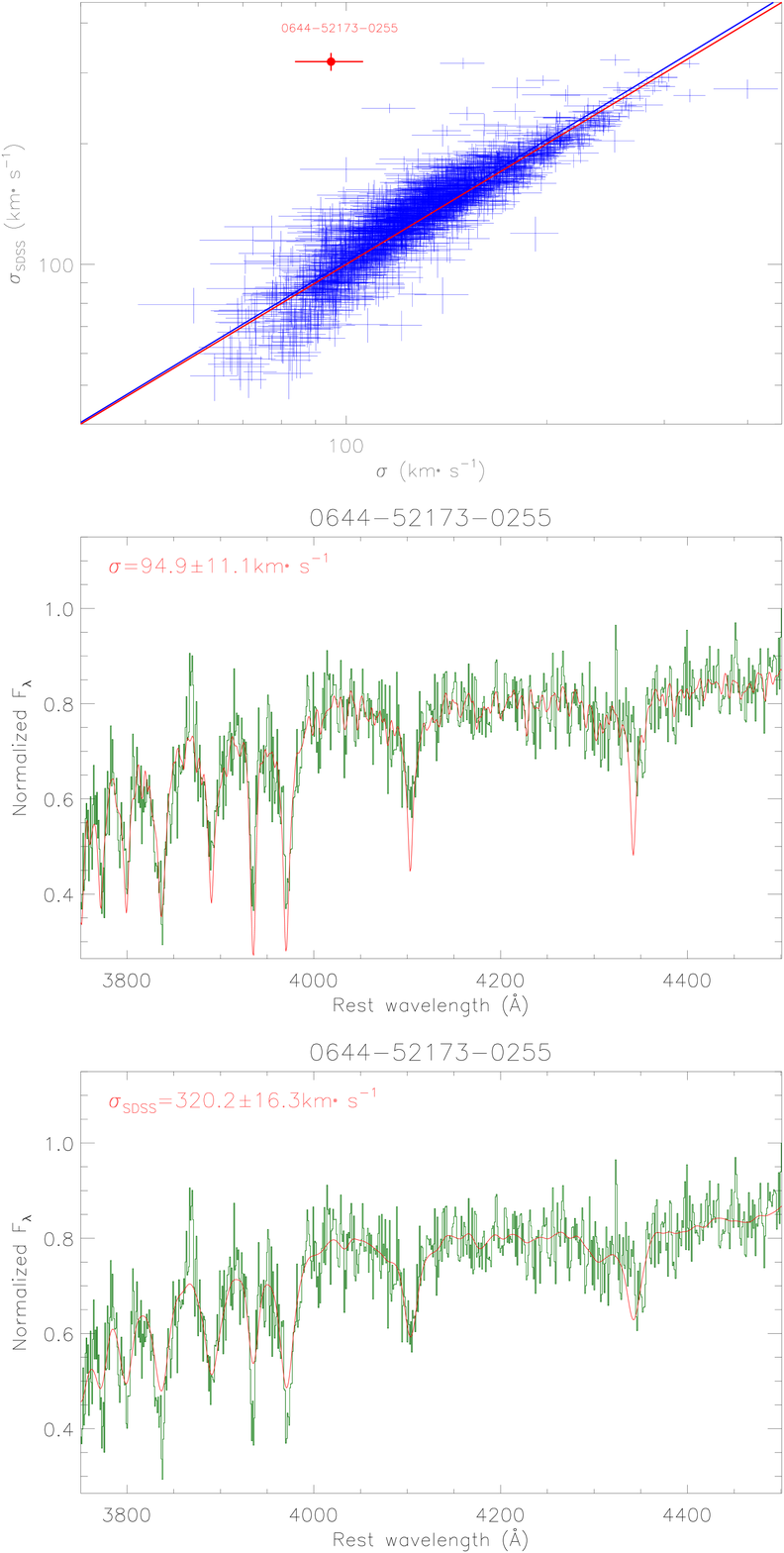}
\caption{Top panel shows the correlation between the measured stellar velocity dispersions 
$\sigma$ and the SDSS pipeline provided $\sigma_{\rm SDSS}$ for the Type-2 AGN. Solid blue 
line shows the best fitting results, and solid red line shows $\sigma~=~\sigma_{\rm SDSS}$. 
Middle and bottom panel show the best fitting results to the absorption features around 
4000\AA~ by the fitting procedure in the manuscript with $\sigma\sim95{\rm km/s}$, and the 
fitting results to the absorption features around 4000\AA~ with stellar velocity dispersion 
fixed to the SDSS pipeline provided $320{\rm km/s}$.
}
\label{sigma2}
\end{figure}

\subsection{Parent samples of Type-1 AGN and Type-2 AGN}
	The work is starting from parent samples of Type-1 AGN and Type-2 AGN. Here, only one 
criterion of redshift smaller than 0.3 ($z~<~0.3$) is applied to collect all the low redshift 
Type-1 AGN from SDSS pipeline classified QSOs \citep{rg02, ro12, pr15, lh20} in DR12, through 
the SDSS provided SQL (Structured Query Language) Search tool 
(\url{http://skyserver.sdss.org/dr12/en/tools/search/sql.aspx}) by the following query
\begin{lstlisting}
SELECT plate, fiberid, mjd
FROM SpecObjall
WHERE
   class='QSO' and z<0.30 and zwarning=0
\end{lstlisting}
where 'SpecObjall' is the SDSS pipeline provided database including basic properties of spectroscopic 
emission features of emission line galaxies in SDSS DR12. More detailed information of the database 
'SpecObjall' can be found in \url{http://skyserver.sdss.org/dr12/en/help/docs/tabledesc.aspx}. The SQL 
query leads 12342 QSOs collected as Type-1 AGN included in the parent sample of Type-1 AGN. And the 
collected information of plate, fiberid and mjd can be conveniently applied to download SDSS spectra 
of the 12342 Type-1 AGN.

	The same criteria $z~<~0.3$ combining with criterion of 'subclass='AGN'' are applied to collect 
all the low redshift Type-2 AGN from SDSS pipeline classified main galaxies in DR12 by the following query
\begin{lstlisting}
SELECT plate, fiberid, mjd
FROM SpecObjall
WHERE
   class='GALAXY' and z<0.30 and zwarning=0
   and subclass = 'AGN'
\end{lstlisting}
where 'subclass='AGN'' can be applied to ensure the collected narrow emission line objects well classified 
as AGN based on the emission line properties. More detailed information of SDSS spectroscopic catalogs 
can be found in \url{https://www.sdss.org/dr12/spectro/catalogs/}. The query above leads 16269 narrow 
emission line galaxies collected as Type-2 AGN included in the parent sample of Type-2 AGN. Therefore, 
in the parent samples, there are 12342 Type-1 AGN and 16269 Type-2 AGN.

	In the manuscript, high redshift ($z~>~0.3$) SDSS AGN are not considered, not only because 
that it is hard to measure reliable stellar velocity dispersions $\sigma$ through spectroscopic features 
of high redshift SDSS AGN, but also because that the SDSS spectra of high redshift AGN will not completely 
cover the broad H$\alpha$ emission features which will be applied to determine whether the collected 
Type-1 AGN have apparent broad emission lines.

\subsection{Method to measure the stellar velocity dispersions}

	Stellar velocity dispersions are measured through absorption features within rest wavelength 
from 3750 to 4600\AA~ described by one broadened template stellar spectrum plus one four-degree 
polynomial function\footnote{Higher order polynomial functions have few effects on the final measured 
velocity dispersions.}, similar as the direct fitting method proposed and discussed in \citet{rw92, 
ba02, gh06}. Meanwhile, when the absorption features are described, the narrow emission lines are 
being masked out by line width (full width at zero intensity) about 450${\rm km~s^{-1}}$, including 
H$\theta$, $H\eta$, He~{\sc i}$\lambda3889$\AA, Ca~H \& K, [S~{\sc ii}]$\lambda4072$\AA, H$\delta$, 
H$\gamma$ and [O~{\sc iii}]$\lambda4364$\AA, the broad emission lines are being masked out by full 
width at zero intensity about 3000${\rm km~s^{-1}}$, especially for the Balmer lines. Here, the 1273 
template stellar spectra with high resolution about ${\rm 30~km\cdot~s^{-1}}$ are collected from 
Indo-U.S. Coude Feed Spectral Library\footnote{\url{https://www.noao.edu/cflib/}} \citep{vg04}. Then, 
through Levenberg-Marquardt least-squares minimization technique\footnote{
\url{https://pages.physics.wisc.edu/~craigm/idl/cmpfit.html}}, absorption features around 4000\AA~ can 
be best described by the most appropriate single template stellar spectrum, leading to the well measured 
stellar velocity dispersions $\sigma$ and corresponding uncertainties. Because of stronger AGN continuum 
emissions in SDSS spectra of Type-1 AGN, stellar velocity dispersions can be well measured in about 
half of the collected Type-1 AGN, but in almost all the Type-2 AGN. Here, we do not show the best 
fitting results to the absorption features around 4000\AA~ in all the AGN, but Fig.~\ref{sigma} shows 
two examples, one Type-1 AGN and one Type-2 AGN, on the best fitting results to the absorption features 
around 4000\AA. Among the several methods to measured stellar velocity dispersions (such as the following 
discussed Simple Stellar Population method), the direct fitting method in the manuscript is mainly 
considered due to the following reason. Among the Type-1 AGN, absorption features of host galaxies 
are not apparent in the whole spectra but apparent enough around 4000\AA, the direct fitting method 
applied to the absorption features around 4000\AA~ can lead as many Type-1 AGN as possible with stellar 
velocity dispersions well measured.

	Rather than Ca~{\sc ii}$\lambda8498,~8542,~8662$\AA~ triplets and Mg~{\sc i}b absorption 
features around 5175\AA, the absorption features around 4000\AA~ are mainly considered in the 
manuscript, due to the following main reasons. On the one hand, for Type-1 AGN in SDSS, only dozens 
of objects with redshift less than 0.06 have high quality Ca~{\sc ii} triplets, leading to a quite small 
sample of Type-1 AGN of which stellar velocity dispersions can be well measured through Ca~{\sc ii} 
triplets. Moreover, in the following section, it will be confirmed that the measured stellar velocity 
dispersions are consistent through the Ca~{\sc ii} triplets and through the absorption features around 
4000\AA. On the other hand, not similar as Ca~{\sc ii} triplets and absorption features around 4000\AA, 
Mg~{\sc i}b absorption features in Type-1 AGN have serious contaminations from optical Fe~{\sc ii} 
emissions and prominent Fe~{\sc i} blends. \citet{gh06} have clearly pointed out that the absorption 
features of the Ca H+K region may provide the only recourse for estimating stellar velocity dispersions, 
at higher AGN contaminations. Similar discussions can also be found in \citet{hb12} which have shown 
that using Ca H\&K instead of CaT seems to be working relatively well.

	Before proceeding further, it is necessary to ensure broad Balmer emission lines in SDSS 
spectra of the collected Type-1 AGN but no broad lines in the collected Type-2 AGN. Therefore emission 
line properties should be further checked after subtractions of host galaxy contributions (if there 
are). The commonly applied SSP method (Simple Stellar Population) has been applied to determine 
contributions of host galaxies. The method above to measured stellar velocity dispersions through a 
single stellar template spectrum is not appropriate to determine host galaxy contributions. More 
detailed descriptions on the SSP method can be found in \citet{bc03, ka03, cm05, cm17} and in our 
previous paper \citet{zh14, zh16, ra17, zh19, zh21m, zh21a, zh21b}. Here, we do not show further 
detailed discussions on the SSP method any more, but simple descriptions on SSP method as follows. 
The 39 simple stellar population templates from \citet{bc03} have been exploited, which can be used 
to well-describe the characteristics of almost all the SDSS galaxies as detailed discussions in 
\citet{bc03}. Meanwhile, there is an additional component, a power law component, which is applied 
to describe intrinsic AGN continuum emissions, especially when the SSP method is applied to describe 
spectra of Type-1 AGN. And the power law component is not limited to be blue, because intrinsic host 
galaxy reddening effects can lead to red power law component, such as the following shown results in 
the top left panel of Fig.~\ref{line} in the Type-1 AGN SDSS 0997-52734-0303. Meanwhile, when the 
SSP method is applied, the narrow emission lines listed in 
\url{http://classic.sdss.org/dr1/algorithms/speclinefits.html#linelist} are masked out by full width 
at zero intensity about 450${\rm km~s^{-1}}$, And the wavelength ranges from 4450 to 5600\AA~ and 
from 6250 to 6750\AA~ are also masked out for the probably broad H$\beta$ and the broad H$\alpha$ 
emission lines. Then, through the Levenberg-Marquardt least-squares minimization technique, SDSS 
spectra with emission lines being masked out can be well described. Here, the SSP method determined 
host galaxy contributions are not shown in plots for all the AGN, but for the two AGN shown in 
Fig.~\ref{sigma}, the SSP determined host galaxy contributions are shown in left panels of Fig.~\ref{line}.

	After subtractions of host galaxy contributions (if there are), emission lines around H$\alpha$ 
within rest wavelength from 6250 to 6850\AA~ can be carefully measured, in order to check whether 
there are broad H$\alpha$ emission lines. To measure properties of emission lines are not the objective 
of the manuscript, but simple descriptions are as follows on the emission line fitting procedure. 
Similar as what we have recently done in \citet{zh21a, zh21b, zh21c}, three broad Gaussian functions 
(second moment larger than 600${\rm km\cdot s^{-2}}$) are applied to describe broad H$\alpha$, 
and seven narrow Gaussian components (second moment smaller than 600${\rm km\cdot s^{-2}}$) are 
applied to describe narrow H$\alpha$, [O~{\sc i}], [N~{\sc ii}] and [S~{\sc ii}] doublets, and a 
power law component is applied to describe continuum emissions underneath broad H$\alpha$. 
Based on the measured parameters of broad H$\alpha$ through the Levenberg-Marquardt least-squares 
minimization technique, the criteria are accepted to determine that there are no broad H$\alpha$: 
the determined three broad Gaussian components for broad H$\alpha$ with the measured line fluxes and 
line widths 2 times smaller than the corresponding uncertainties, and the criteria are accept to 
determine that there are reliable broad H$\alpha$: there are at least one broad Gaussian component 
with measured line flux and line width at least 5 times larger than their corresponding uncertainties 
and second moment larger than 600${\rm km\cdot~s^{-1}}$. The best descriptions to the emission lines 
around H$\alpha$ are not shown in plots for all the AGN, but for the two AGN shown in Fig.~\ref{sigma}, 
the best descriptions to the emission lines around H$\alpha$ are shown in right panels of Fig.~\ref{line}.

	Furthermore, emission line properties of [O~{\sc iii}] line should be applied in the manuscript. 
Therefore, simple descriptions are shown on the model functions to describe emission lines around 
H$\beta$ within rest wavelength from 4400 to 5600\AA~ after subtractions of host galaxy contributions. 
Similar as what we have recently done in \citet{zh21a, zh21b, zh21c}, three broad Gaussian functions 
are applied to describe broad H$\beta$, one narrow Gaussian component is applied to describe narrow 
H$\beta$, two narrow and two broad Gaussian components are applied to describe core and extended 
components of [O~{\sc iii}]$\lambda4959,~5007$\AA~ doublet, one Gaussian component is applied to describe 
He~{\sc ii} line, broadened and scaled Fe~{\sc ii} templates discussed in \citet{kp10} are applied to 
describe probable optical Fe~{\sc ii} lines, and a power law component is applied to describe continuum 
emissions underneath broad H$\beta$. Fig.~\ref{hb} shows two examples on the best-fitting results to 
emission lines around H$\beta$ through the Levenberg-Marquardt least-squares minimization technique.

	Finally, based on the following criteria, 
\begin{itemize}
\item The measured stellar velocity dispersions are at least 5 times larger than their 
	corresponding uncertainties.
\item The measured stellar velocity dispersions are larger than 50~${\rm km\cdot~s^{-1}}$ and 
	smaller than 400~${\rm km\cdot~s^{-1}}$.
\item The $\chi^2$ should be smaller than 2, based on the best fitting results to the absorption 
	features around 4000\AA.
\item For the Type-1 AGN, there are reliable broad emission lines with at least one broad Gaussian 
	component with the measured line flux and line width at least 5 times larger than the 
	corresponding uncertainties and second moment larger than 600${\rm km\cdot~s^{-1}}$.
\item For the Type-2 AGN, there are no broad emission lines with the determined three broad 
	Gaussian components for broad H$\alpha$ with the measured line fluxes and line widths 2 
	times smaller than the corresponding uncertainties.
\end{itemize} 
there are 6260 Type-1 AGN with apparent broad H$\alpha$ emission lines and reliable stellar 
velocity dispersions, and 15353 Type-2 AGN with reliable stellar velocity dispersions but no 
broad H$\alpha$ emission lines. And the sample of the 6260 Type-1 AGN is so far the largest 
sample of Type-1 AGN with reliable measured stellar velocity dispersions, about 85 times larger 
than the more recent sample of Type-1 AGN with measured stellar velocity dispersions in 
\citet{bt15, bt21}. And through absorption features around 4000\AA, there are 50.7\% (6260 of 
12342) of Type-1 AGN with stellar velocity dispersions well measured, and 94.4\% (15353 of 
16269) of Type-2 AGN with stellar velocity dispersions well measured.

	The necessary parameters of all the 6260 Type-1 AGN and all the 15353 Type-2 AGN are not 
listed in the manuscript, but can be downloaded from \url{https://pan.baidu.com/s/1NCDqFtJwRaG-u21ekxlvvQ} 
with validation code sj6f. There are 11 columns in the data file of type1\_vd.list and type2\_vd.list 
in ASCII format for the 6260 Type-1 AGN and the 15353 Type-2 AGN with information of SDSS PLATE-MJD-FIBERID, 
measured $\sigma$ and uncertainty in unit of ${\rm km\cdot~s^{-1}}$ without corrections of instrument 
resolutions, $\chi^2$ value for the best fitting results to the absorption features around 4000\AA, 
redshift, logarithmic line luminosity and uncertainty of total [O~{\sc iii}] line in unit of 
${\rm erg\cdot~s^{-1}}$, logarithmic line luminosity and uncertainty of core component of [O~{\sc iii}] 
line in unit of ${\rm erg\cdot~s^{-1}}$, logarithmic ratio of O3HB (flux ratio of core component of 
[O~{\sc iii}]$\lambda5007$\AA~ to narrow H$\beta$), logarithmic ratio of N2HA (flux ratio of 
[N~{\sc ii}]$\lambda6583$\AA~ to narrow H$\alpha$).

\section{To confirmed few effects of AGN activities on the measured stellar velocity dispersions 
in Type-1 AGN}

	Before proceeding further, it is necessary to check effects of AGN activities, especially 
power law AGN continuum emissions and broad line emissions, on our measured stellar velocity dispersions 
through absorption features around 4000\AA~ by the fitting procedure described in the Section above, 
due to apparent AGN continuum emissions and broad line emissions in Type-1 AGN. In the section, series 
of artificial spectra are created including different contributions of AGN continuum emissions and 
broad emission lines, and then to check whether are there apparent effects of continuum emissions or 
broad line emissions on the measured stellar velocity dispersions in Type-1 AGN.

	Based on SDSS spectra $S_2(\lambda)$ of Type-2 AGN with well measured stellar velocity 
dispersions, artificial spectra $S_{a}(\lambda)$ can be created with contributions of AGN continuum 
emissions $P_{AGN}(\lambda)$ and broad line emissions $L_{br}(\lambda)$, 
\begin{equation}
\begin{split}
	S_{a}(\lambda)~&=~S_2(\lambda)~+~P_{AGN}(\lambda)~+~L_{br}(\lambda) \\
	&=~S_2(\lambda)~+~C_0\times(\frac{\lambda}{5100\textsc{\AA}})^{\alpha}\\
	\ \ \ \ \ &~+~\sum_{i=1}^{3}Gauss1(\lambda, [\lambda_{0,i}, \sigma_{0}, f_{0,i}])
\end{split}
\end{equation}
where $P_{AGN}(\lambda)=C_0\times(\frac{\lambda}{5100\textsc{\AA}})^{\alpha}$ means AGN continuum 
emissions, $L_{br}(\lambda)~=~\sum_{i=1}^{3}Gauss1(\lambda, [\lambda_{0,i}, \sigma_{0,i}, f_{0,i}])$ 
means Gaussian like broad emission lines. For AGN continuum emissions, $\alpha$ is randomly selected 
from -2 to 0.5, as well discussed results on composite spectrum of SDSS QSOs in \citet{van01}. And 
for broad emission lines within rest wavelength range from 3750\AA~ to 4600\AA, the three broad 
lines of H$\delta$ ($\lambda_{0,1}=4103$\AA), H$\gamma$ ($\lambda_{0,2}=4341$\AA) and the broad 
optical Fe~{\sc ii} (opt37, opt38, $\lambda_{0,3}\sim4565$\AA) are mainly considered. The other broad 
line features (such as H$\epsilon$ and the other optical Fe~{\sc ii} features) within rest wavelength 
from 3750 to 4600\AA~ are not considered, because they are quite weak. Flux ratios of the three 
broad line features ($f_{0,1}$ as line flux of broad H$\delta$, $f_{0,2}$ as line flux of broad 
H$\gamma$, $f_{0,3}$ as line flux of the optical Fe~{\sc ii} feature) are accepted as 
$f_{0,1}:f_{0,2}:f_{0:3}=5.05:12.62:3.76$ from the composite spectrum of SDSS QSOs. The same second 
moments from 600 to 2000$km/s$ are accepted to the three broad line features. Accepted the strong 
correlation between continuum luminosity at 5100\AA~ and the H$\beta$ luminosity in \citet{gh05b} 
and based on the reported continuum luminosity and broad H$\beta$ luminosity of the quasars in 
SDSS DR7 in \citet{sh11}, the input values of $C_0$ and $f_{0,2}$ are tied to be 
\begin{equation}
\frac{3.3~\times~f_{0,2}}{10^{-17}{\rm erg/s/cm^2}}~\sim~51\times
	\frac{C_0}{10^{-17}{\rm erg/s/cm^2/\textsc{\AA}}}
\end{equation}
where the factor 3.3 is the flux ratio of H$\beta$ to H$\gamma$.

	Based on randomly selected $C_0$ from 0.2 to 10 times of the mean intensity of $S_2(\lambda)$, 
$\alpha$ randomly selected from -2 to 0.5 and $\sigma_{0}$ randomly selected from 600 to 2000$km/s$ 
and randomly collected $S_2(\lambda)$ among the 11353 Type-2 AGN, 1000 artificial spectra are 
created with contributions of both AGN continuum emissions and broad line emissions. If there were 
apparent effects of AGN activities on the measured stellar velocity dispersions, re-measured stellar 
velocity dispersions $\sigma_a$ in $S_{a}(\lambda)$ should be quite different from the stellar velocity 
dispersions $\sigma_i$ in $S_2(\lambda)$ of Type-2 AGN. Left panel of Fig.~\ref{sax} shows an example 
of $S_{a}(\lambda)$ and the best fitting results to the absorption features in the artificial spectrum. 
Right panel of Fig.~\ref{sax} shows the correlation between artificial $\sigma_a$ and intrinsic 
$\sigma_i$ of the 628 artificial spectra with measured $\sigma_a$ five times larger than corresponding 
uncertainties. There are no reliable measured stellar velocity dispersions in the other 372 artificial 
spectra which have the mean $C_0$ about two times higher than the $C_0$ applied in the 628 artificial 
spectra with reliable measured stellar velocity dispersions, due to their measured $\sigma_a$ three 
times smaller than their corresponding uncertainties. There is a strong positive linear correlation 
with Spearman Rank correlation coefficient about 0.92 with $P_{null}~<~10^{-10}$ for the results shown 
in right panel of Fig.~\ref{sax}. Under considerations of uncertainties in both coordinates, the 
correlation can be described by
\begin{equation}
\log(\frac{\sigma_{i}}{\rm km/s})~=~(-0.046\pm0.047)~+~
	(0.976\pm0.022)\times\log(\frac{\sigma_a}{\rm km/s})
\end{equation}
through the FITEXY code (\url{https://idlastro.gsfc.nasa.gov/ftp/pro/math/fitexy.pro} written by 
Frank Varosi) \citep{tr02}. Meanwhile, the mean ratio of artificial $\sigma_a$ to intrinsic $\sigma_i$ 
is about 1.04$\pm$0.03, with the uncertainty 0.03 estimated through the bootstrap method with 1000 
loops. For each loop, a new sample of $\sigma_{a,pos}/\sigma_{i,pos}$ is created with more than half 
data points randomly collected from the sample of $\sigma_a/\sigma_i$. After 1000 loops, there are 
1000 new samples with 1000 mean values, the half width at half maximum of distribution of the 1000 
mean values is accepted as the uncertainty of the mean value of $\sigma_a/\sigma_i$.

	Moreover, the mean ratio 1.04 of artificial $\sigma_a$ to intrinsic $\sigma_i$ kindly 
larger than 1 apparently indicates central AGN continuum emissions and/or broad emission lines 
can lead the measured stellar velocity dispersions to be about 4\% larger than the intrinsic values 
in Type-1 AGN. In order to confirm the mean ratio of artificial $\sigma_a$ to intrinsic $\sigma_i$, 
one another sample of 1000 artificial spectra are created. The totally same mean ratio 1.04 can be 
found in the new sample of 1000 artificial spectra. Therefore, the mean ratio of artificial $\sigma_a$ 
to intrinsic $\sigma_{i}$ about 1.04 is intrinsically true, due to effects of central AGN continuum 
emissions and/or broad emission lines.

	Before the end of the section, one point is noted. In right panel of Fig.~\ref{sax}, there 
are some outliers. However, there is only one criterion collect artificial $\sigma_a$ that $\sigma_a$ 
at least 5 times larger than their corresponding uncertainties. If firm criteria are applied that 
$\chi^2$ smaller than 2 (the critical value applied in Section 2) and $\sigma_a$ at least 10 times 
larger than their corresponding uncertainties, the collected 262 data points marked as red symbols 
are well lying within 99.99\% confidence bands of the linear correlation $Y=X$. Moreover, the mean 
ratio of artificial $\sigma_a$ to intrinsic $\sigma_i$ of the 262 high quality data points is about 
1.038 totally similar as the 1.04 for all the data points. Therefore, the outliers are only due to 
rough selection criteria, and there are no further discussions on the outliers in the Fig.~\ref{sax} 
which have few effects on our final results.

	Contributions of AGN continuum emissions and broad line emissions have apparent effects on 
measuring stellar velocity dispersions in the Type-1 AGN, the effects can lead to unmeasured stellar 
velocity dispersions in the Type-1 AGN with strong AGN continuum emissions, meanwhile, the effects 
can lead the measured reliable stellar velocity dispersions in the Type-1 AGN to be about 4\% larger 
than their intrinsic values, which will be carefully discussed in the following stellar velocity 
dispersion comparisons between Type-2 AGN and Type-1 AGN.

\section{To confirm the reliability of the measured stellar velocity dispersions}

	The main objective of the section is to provide evidence to confirm/support the reliability 
of our measured stellar velocity dispersions in Type-2 AGN and in Type-1 AGN, by comparing our 
measured values and reported values in the literature.

\begin{figure*}  
\centering\includegraphics[width = 18cm,height=20cm]{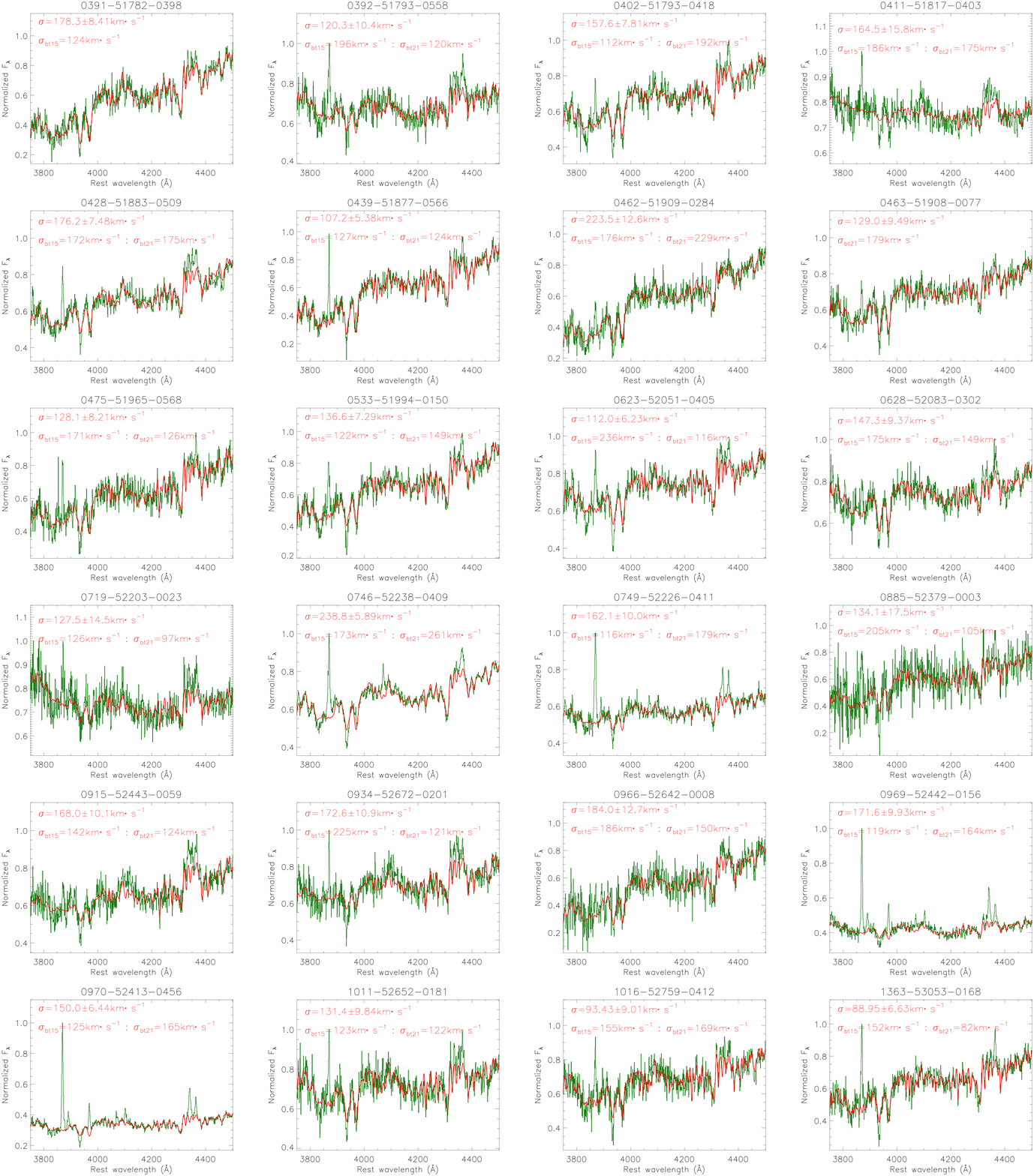}
\caption{Best fitting results (solid red line) to the absorption features around 4000\AA~ (solid 
dark green line) of the 41 Type-1 AGN included in the sample of \citet{bt15} and in the sample 
of \citet{bt21}, and the 2 Type-1 AGN from the sample of \citet{xb11} and 5 Type-1 AGN from the 
sample of \citet{hb12}. The measured stellar velocity dispersion $\sigma$ in the manuscript after 
corrections of instrument resolutions and the reported value $\sigma_{bt15}$ in \citet{bt15} and/or 
$\sigma_{bt21}$ in \citet{bt21}, or $\sigma_{xb11}$ in \citet{xb11}, or $\sigma_{hb12}$ in 
\citet{hb12} are marked in red characters in each panel. 
}
\label{bt15}
\end{figure*}

\setcounter{figure}{5}

\begin{figure*}
\centering\includegraphics[width = 18cm,height=20cm]{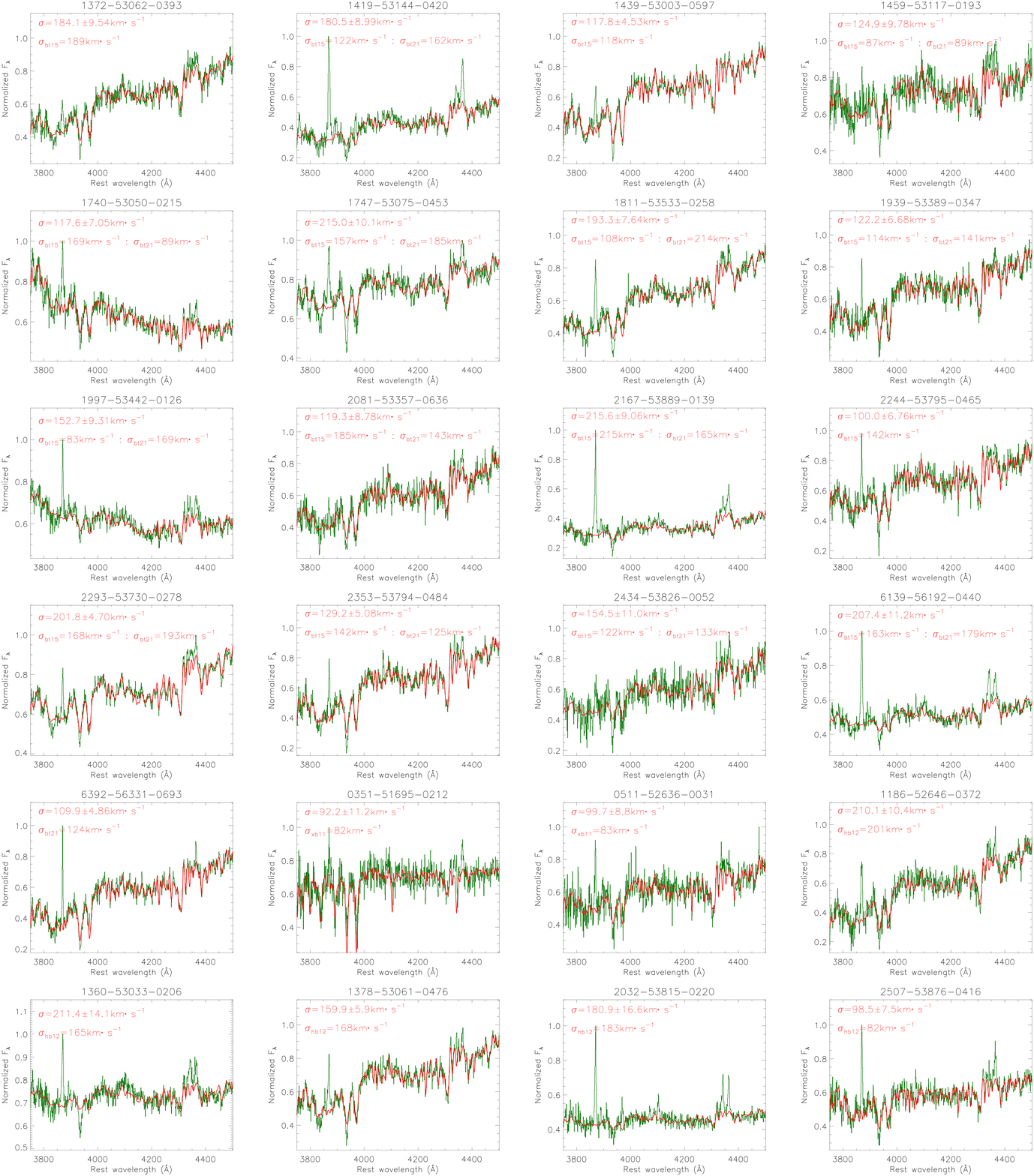}
\caption{--to be continued}
\end{figure*}

	For Type-2 AGN, the measured $\sigma$ are compared with SDSS provided values $\sigma_{\rm SDSS}$, 
shown in top panel of Fig.~\ref{sigma2}. The SDSS pipeline determined $\sigma_{\rm SDSS}$ are based on 
the 24 broadened template eigen-spectra from the ELODIE stellar library well discussed in \citet{ps01} 
applied to describe the whole host galaxy absorption features in SDSS spectra, through the similar minimum 
least-squares minimization technique. More discussions on the SDSS pipeline determined $\sigma_{\rm SDSS}$ 
can be found in \citet{sh12, ts13}. More detailed information of the SDSS pipeline determined 
$\sigma_{\rm SDSS}$ can be found in \url{https://www.sdss.org/dr12/algorithms/redshifts/}. More recently, 
\citet{wo16} have compared the SDSS pipeline provided stellar velocity dispersions with the [O~{\sc iii}] 
line width on studying properties of outflows in Type-2 AGN.

	Between the measured $\sigma$ and the $\sigma_{\rm SDSS}$, there is one strong linear correlation 
with Spearman Rank correlation coefficient of 0.92 with $P_{null}~<~10^{-10}$. Under considerations of 
uncertainties in both coordinates, the correlation can be described by
\begin{equation}
\log(\frac{\sigma_{\rm SDSS}}{\rm km\cdot~s^{-1}})~=~(-0.01\pm0.01)~+~
	(1.01\pm0.05)\times\log(\frac{\sigma}{\rm km\cdot~s^{-1}}) 
\end{equation}
through the FITEXY code. The mean ratio of $\sigma$ to $\sigma_{\rm SDSS}$ is about 0.985$\pm$0.003, 
with the uncertainty 0.003 estimated by the bootstrap method with 1000 loops. The strong linear 
correlation and the mean ratio of $\sigma$ to $\sigma_{\rm SDSS}$ near to 1 clearly indicate that the 
measured $\sigma$ through absorption features around 4000\AA~ are reliable enough for the Type-2 AGN.

	Although there is a strong linear correlation in Fig.~\ref{sigma2}, there are some outliers 
in the space of $\sigma$ versus $\sigma_{\rm SDSS}$, such as the extreme outlier SDSS 0644-52173-0255 
which has $\sigma\sim95{\rm km/s}$ measured through absorption features around 4000\AA~ in the manuscript 
however $\sigma_{SDSS}\sim320{\rm km/s}$ in SDSS database. The best fitting results to the absorption 
features around 4000\AA~ are shown in middle panel of Fig.~\ref{sigma2} with $\chi^2\sim1.02$. And in 
order to show further clues to support our measured stellar velocity dispersion in SDSS 0644-52173-0255, 
bottom panel of Fig.~\ref{sigma2} shows the fitting results to the absorption features around 4000\AA~ 
with stellar velocity dispersion fixed to the SDSS pipeline provided $\sigma_{SDSS}\sim320{\rm km/s}$, 
leading to bad fitting results to the absorption features with rest wavelength from 3800\AA~ to 4000\AA, 
indicating stellar velocity dispersion around $320{\rm km/s}$ not preferred in SDSS 0644-52173-0255. We 
do not know the exact reason leading to so large SDSS pipeline determined stellar velocity dispersion 
in SDSS 0644-52173-0255, but probably due to the following main reason that the SDSS pipeline provided 
stellar velocity dispersion is measured through the whole host galaxy absorption features including 
more contaminations. We do not have an idea to determine how many outliers in the space of $\sigma$ 
versus $\sigma_{\rm SDSS}$, however the strong linear correlation and the mean ratio of $\sigma$ to 
$\sigma_{\rm SDSS}$ being well near to 1 can be applied as strong statistical evidence to support the 
reliability of our measured stellar velocity dispersions in the Type-2 AGN.

	There are no $\sigma_{\rm SDSS}$ provided by SDSS for Type-1 AGN, but the following three 
methods are applied to confirm the reliability of the measured $\sigma$ in Type-1 AGN. First, it 
can be applied to confirm the reliability of our measured stellar velocity dispersions in Type-1 
AGN, by comparing our measured values and the more confident spatially resolved values in \citet{bt15, 
bt21}. The spatially-resolved stellar velocity dispersion measurements from Keck long-slit spectra 
in \citet{bt15, bt21} have more advantages in the measurements than relying only on fiber-based 
SDSS spectra. There are 65 Type-1 AGN included in the sample of \citet{bt15} with measurements 
of stellar velocity dispersions derived for both aperture and spatially resolved spectra, and 66 
Type-1 AGN in the sample of \citet{bt21} (63 Type-1 AGN included in the sample of \citet{bt15} and 
3 new Type-1 AGN) with determined stellar velocity dispersions from spatially resolved measurements 
integrated within effective spheroid radius. Among the 65 Type-1 AGN in \citet{bt15} and the 66 
Type-1 AGN in \citet{bt21}, there are 41 AGN of which $\sigma$ can be well measured through absorption 
features around 4000\AA. Among the 41 Type-1 AGN, 2 Type-1 AGN is firstly reported in the sample 
of \citet{bt21}, 4 Type-1 AGN are only reported in the sample of \citet{bt15}, the other 35 Type-1 
AGN are reported both in the sample of \citet{bt15} and in the sample of \citet{bt21}. The best 
fitting results to the absorption features around 4000\AA~ are shown in Fig.~\ref{bt15} for the 
41 Type-1 AGN, with the measured stellar velocity dispersions $\sigma$ in the manuscript and the 
reported stellar velocity dispersions $\sigma_{bt15}$ in \citet{bt15} and $\sigma_{bt21}$ in 
\citet{bt21} marked in each panel of Fig.~\ref{bt15} (the velocity dispersions are not listed in a 
table any more).

\begin{figure}
\centering\includegraphics[width = 8cm,height=5cm]{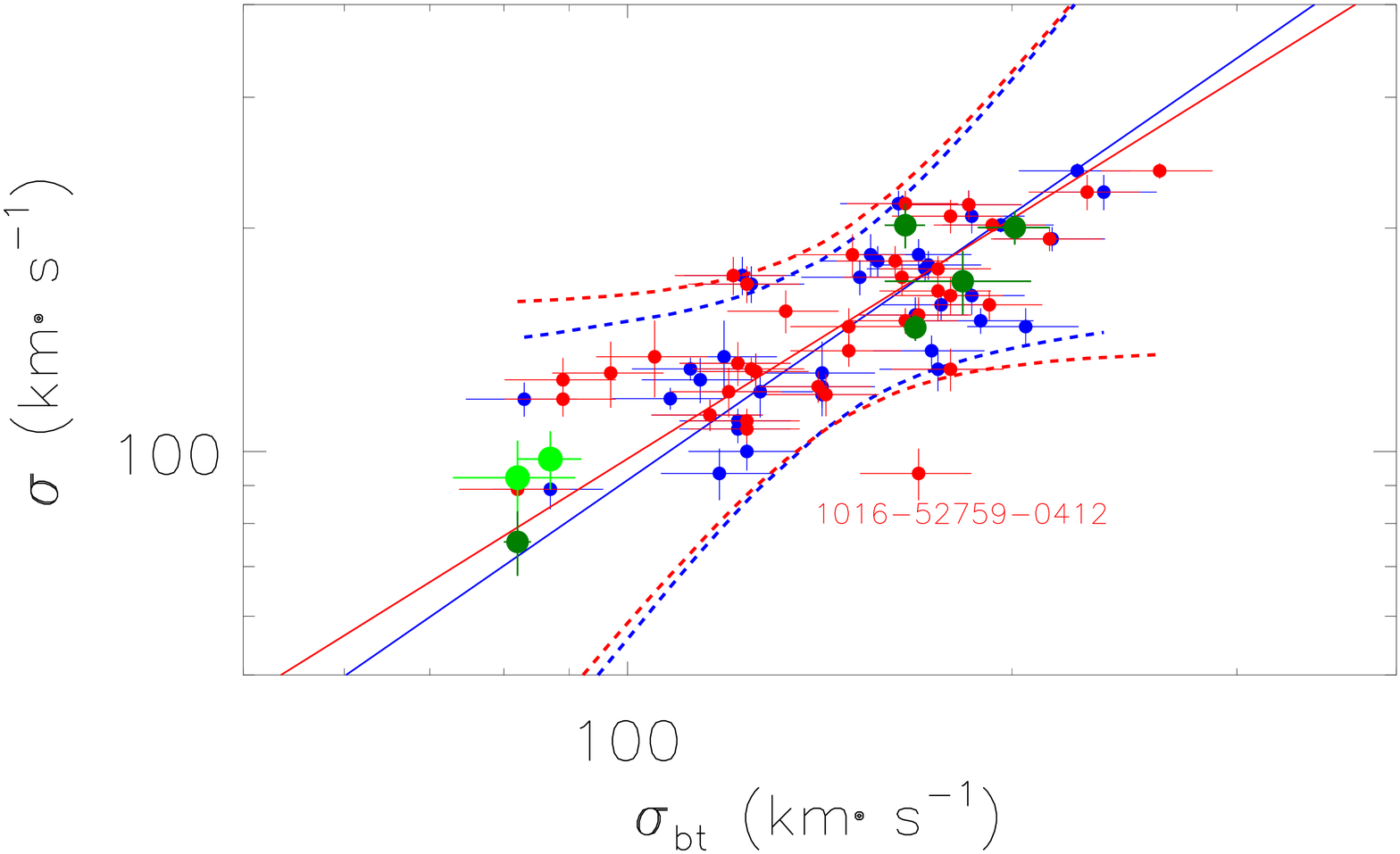}
\caption{Correlations between the measured stellar velocity dispersions $\sigma$ after corrections 
of instrument resolutions and the reported $\sigma_{bt15}$ (in blue color) of the 39 Type-1 AGN in 
\citet{bt15}, between $\sigma$ and $\sigma_{bt21}$ (in red color) of the 37 Type-1 AGN in \citet{bt21}, 
and the reported $\sigma_{xb11}$ (in green color) of the 2 Type-1 AGN in \citet{xb11} and the reported 
$\sigma_{hb12}$ (in dark green color) of the 5 Type-1 AGN in \citet{hb12}. Solid blue line and dashed 
blue lines show the best fitting results to the correlation between $\sigma_{bt15}$ and the measured 
$\sigma$ in the manuscript after corrections of instrument resolutions, and the corresponding 99.9\% 
confidence bands. Solid red line and dashed red lines show the best fitting results to the correlation 
between $\sigma_{bt21}$ and the measured $\sigma$ in the manuscript after corrections of instrument 
resolutions, and the corresponding 99.9\% confidence bands.
}
\label{bt}
\end{figure}

	Then, comparisons are shown in Fig.~\ref{bt} between $\sigma$ and $\sigma_{bt}$, with 
Spearman Rank correlation coefficients of 0.76 with $P_{null}~<~3.6\times10^{-8}$ for the 39 Type-1 
AGN with $\sigma_{bt15}$ in \citet{bt15} and of 0.71 with $P_{null}~<~1.7\times10^{-6}$ for the 37 
Type-1 AGN with $\sigma_{bt21}$ reported in \citet{bt21}. Here, in order to find more reasonable 
results, similar as discussed in \citet{gh05}, effects of SDSS instrument resolutions can be 
corrected on the measured stellar velocity dispersions by 
\begin{equation}
\sigma^2~=~\sigma_{obs}^2~-~\sigma_{in}^2~+~\sigma_{tp}^2
\end{equation}
where $\sigma_{obs}$ represents the measured stellar velocity dispersions, 
$\sigma_{in}~\sim~70{\rm km\cdot~s^{-1}}$ represents the SDSS instrument resolution around 4000\AA, 
and $\sigma_{tp}~\sim~30{\rm km\cdot~s^{-1}}$ represents the instrument resolution of the applied 
stellar template. Under considerations of uncertainties in both coordinates, through the FITEXY code, 
the linear correlations can be described by
\begin{equation}
\begin{split}
\log(\frac{\sigma}{\rm km\cdot~s^{-1}})~&=~(-0.37\pm0.22)~+~ \\
	&\ \ \ \ (1.17\pm0.10)\times\log(\frac{\sigma_{bt15}}{\rm km\cdot~s^{-1}}) \\
\log(\frac{\sigma}{\rm km\cdot~s^{-1}})~&=~(-0.08\pm0.19)~+~\\
	&\ \ \ \  (1.04\pm0.09)\times\log(\frac{\sigma_{bt21}}{\rm km\cdot~s^{-1}})
\end{split}
\end{equation}
for the 39 Type-1 AGN with $\sigma_{bt15}$ in \citet{bt15}, and for the 37 Type-1 AGN with 
$\sigma_{bt21}$ reported in \citet{bt21}, respectively. Moreover, the mean ratio of $\sigma$ to 
$\sigma_{bt15}$ is 1.016$\pm$0.045 for the 39 Type-1 AGN in \citet{bt15}, and the mean ratio of 
$\sigma$ to $\sigma_{bt21}$ is 0.999$\pm$0.052 for the 37 Type-1 AGN in \citet{bt21}. The 
uncertainties of the mean ratios are estimated through the bootstrap method with 1000 loops. 
The strong linear correlations provide evidence to support the reliability of the measured stellar 
velocity dispersions in the Type-1 AGN.

	Moreover, besides the reported Type-1 AGN with spatially resolved stellar velocity 
dispersions in \citet{bt15, bt21}, there are three known large samples of SDSS Type-1 AGN with 
measured stellar velocity dispersions, a sample of 76 Seyfert 1 galaxies in \citet{xb11}, the 
Type-1 AGN in the SDSS reverberation mapping project in \citet{sh15a, sh15b, gt17} (SDSSRM) 
and a sample of low redshift Type-1 AGN in \citet{hb12}. However, \citet{xb11} mainly focused 
on Type-1 AGN with lower stellar velocity dispersions, and there are only 2 Type-1 AGN which have 
reliable stellar velocity dispersions measured through Ca~{\sc ii} triplets in \citet{xb11} 
and also have reliable stellar velocity dispersions measured through absorption features around 
4000\AA~ in the manuscript: SDSS J112526+022039 (plate-mjd-fiberid=0511-52636-0031) and SDSS 
J170246+602818 (plate-mjd-fiberid=0351-51695-0212). And there are 17 SDSSRM AGN with redshift 
smaller than 0.3 and with reported stellar velocity dispersions, however, none of the 17 SDSSRM 
AGN is included in our sample of Type-1 AGN in SDSS DR12 with reliable stellar velocity 
dispersions measured absorption features around 4000\AA. Among the Type-1 AGN in \citet{hb12}, 
there are only 5 Type-1 AGN with stellar velocity dispersions reported in \citet{hb12} but not 
reported in \citet{bt15, bt21}, and also included in our main sample of Type-1 AGN with measured 
stellar velocity dispersions through absorption features around 4000\AA. Therefore, besides the 
41 Type-1 AGN collected from \citet{bt15, bt21}, two additional Type-1 AGN are collected from 
\citet{xb11} and 5 additional Type-1 AGN are collected from \citet{hb12}, of which the best 
fitting results to absorption features around 4000\AA~ are shown in the last 7 panels of 
Fig.~\ref{bt15} and the corresponding properties of stellar velocity dispersions are shown as 
solid green circles and as solid dark green circles in Fig.~\ref{bt}.

	Moreover, as shown in Fig.~\ref{bt15} and Fig.~\ref{bt}, there are quite different 
stellar velocity dispersions in individual Type-1 AGN, such as in SDSS 1016-52759-0412 (SDSS 
J114545+554759) with stellar velocity dispersion 118$km/s$ in \citet{bt15} but with 169$km/s$ 
in \citet{bt21}. Quite different stellar velocity dispersions in individual objects are mainly 
due to different definitions of measuring stellar velocity dispersions. The most expected result 
is that stellar velocity dispersions measured by different definitions lead to totally similar 
\msig measurements, therefore, stellar velocity dispersions both in \citet{bt15} and in \citet{bt21} 
are collected and compared in the manuscript. And, the mean ratios of $\sigma$ to $\sigma_{bt}$ 
well near to 1 strongly indicate that the measured stellar velocity dispersions can not lead 
to statistically larger or smaller stellar velocity dispersions in the Type-1 AGN by the method 
in the manuscript. 

\begin{figure*}  
\centering\includegraphics[width = 18cm,height=20cm]{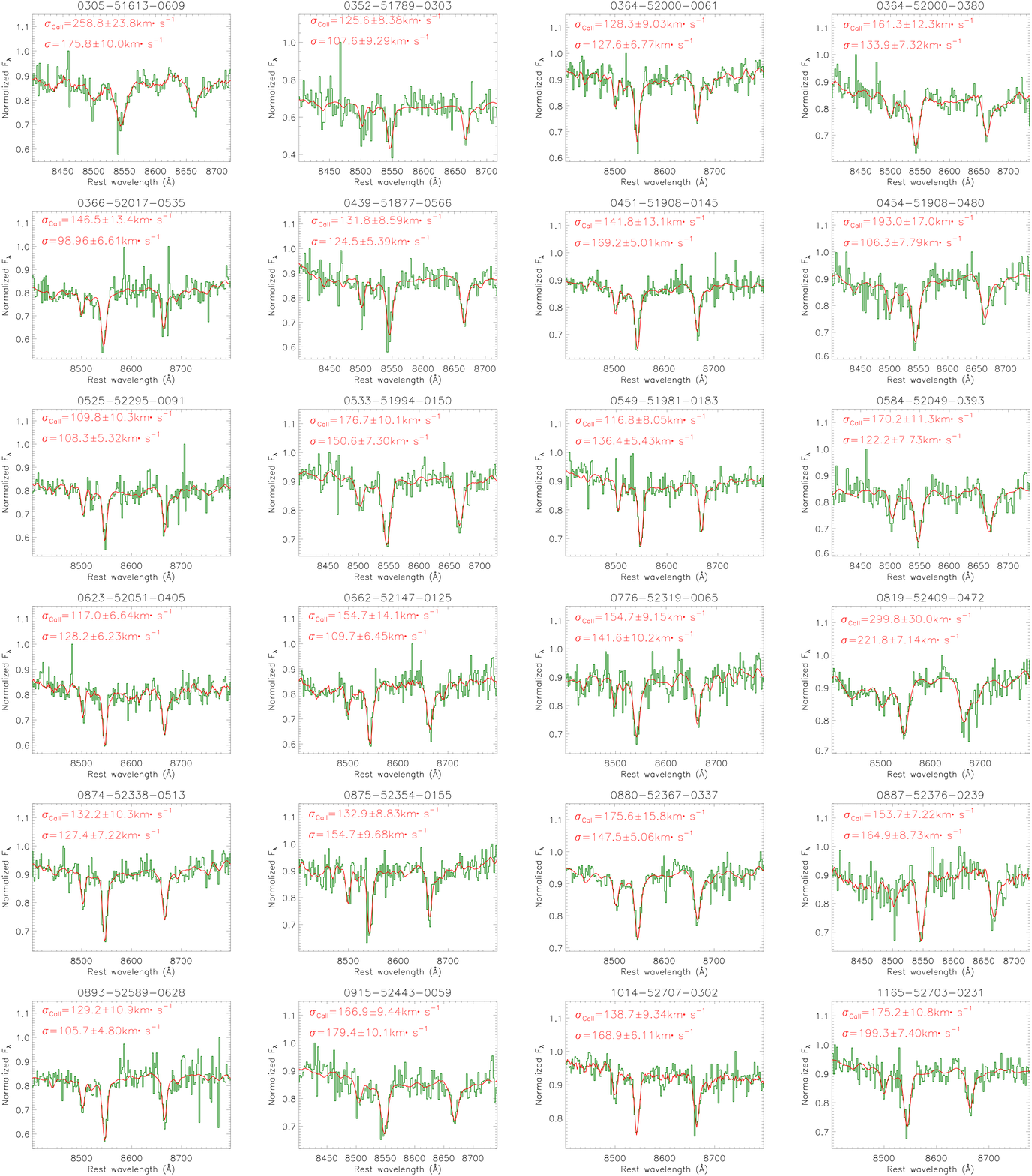}
\caption{Best fitting results (solid red line) to Ca~{\sc ii} triplets (solid dark green line) 
of the 72 Type-1 AGN. The measured stellar velocity dispersion $\sigma_{\rm CaII}$ through 
Ca~{\sc ii} triplets and $\sigma$ through absorption features around 4000\AA~ are marked in 
red characters in each panel.}
\label{caii}
\end{figure*}

	Second, it is interesting to compare the measured $\sigma$ through absorption features 
around 4000\AA~ and dispersions $\sigma_{\rm CaII}$ measured through absorption features of Ca~{\sc ii} 
triplets. Among the Type-1 AGN with measured $\sigma$ through absorption features around 4000\AA, 
there are 72 AGN with high quality Ca~{\sc ii} triplets leading to well measured reliable stellar 
velocity dispersions $\sigma_{\rm CaII}$ (at least 5 times larger than their corresponding 
uncertainties and corresponding $\chi^2$ smaller than 2) in the SDSS spectra with median 
signal-to-noises larger than 20, through the same procedure applied to measure stellar velocity 
dispersions $\sigma$ through absorption features around 4000\AA. Fig.~\ref{caii} shows the best 
fitting results to Ca~{\sc ii} triplets of the 72 Type-1 AGN. The measured $\sigma_{\rm CaII}$ 
and $\sigma$ are marked in each panel, and the measured $\sigma_{\rm CaII}$ are not listed in a 
table any more. Fig.~\ref{sigma1} shows the comparisons between $\sigma_{\rm CaII}$ and $\sigma$, 
through different absorption features, with Spearman Rank correlation coefficient of 0.63 with 
$P_{null}~\sim~4.5\times10^{-9}$. Under considerations of uncertainties in both coordinates, 
through the FITEXY code, the correlation can be described by
\begin{equation}
\log(\frac{\sigma_{\rm CaII}}{\rm km\cdot~s^{-1}})~=~(0.41\pm0.11)~+~
	(0.82\pm0.05)\times\log(\frac{\sigma}{\rm km\cdot~s^{-1}}) 
\end{equation}.
And the mean ratio of $\sigma$ to $\sigma_{CaII}$ is about 0.95$\pm$0.04, with the uncertainty 
calculated by the bootstrap method with 1000 loops. Therefore, based on the results in Fig.~\ref{bt} 
and in Fig.~\ref{sigma1}, the measured $\sigma$ are reliable for the Type-1 AGN.

\setcounter{figure}{7}
\begin{figure*}
\centering\includegraphics[width = 18cm,height=20cm]{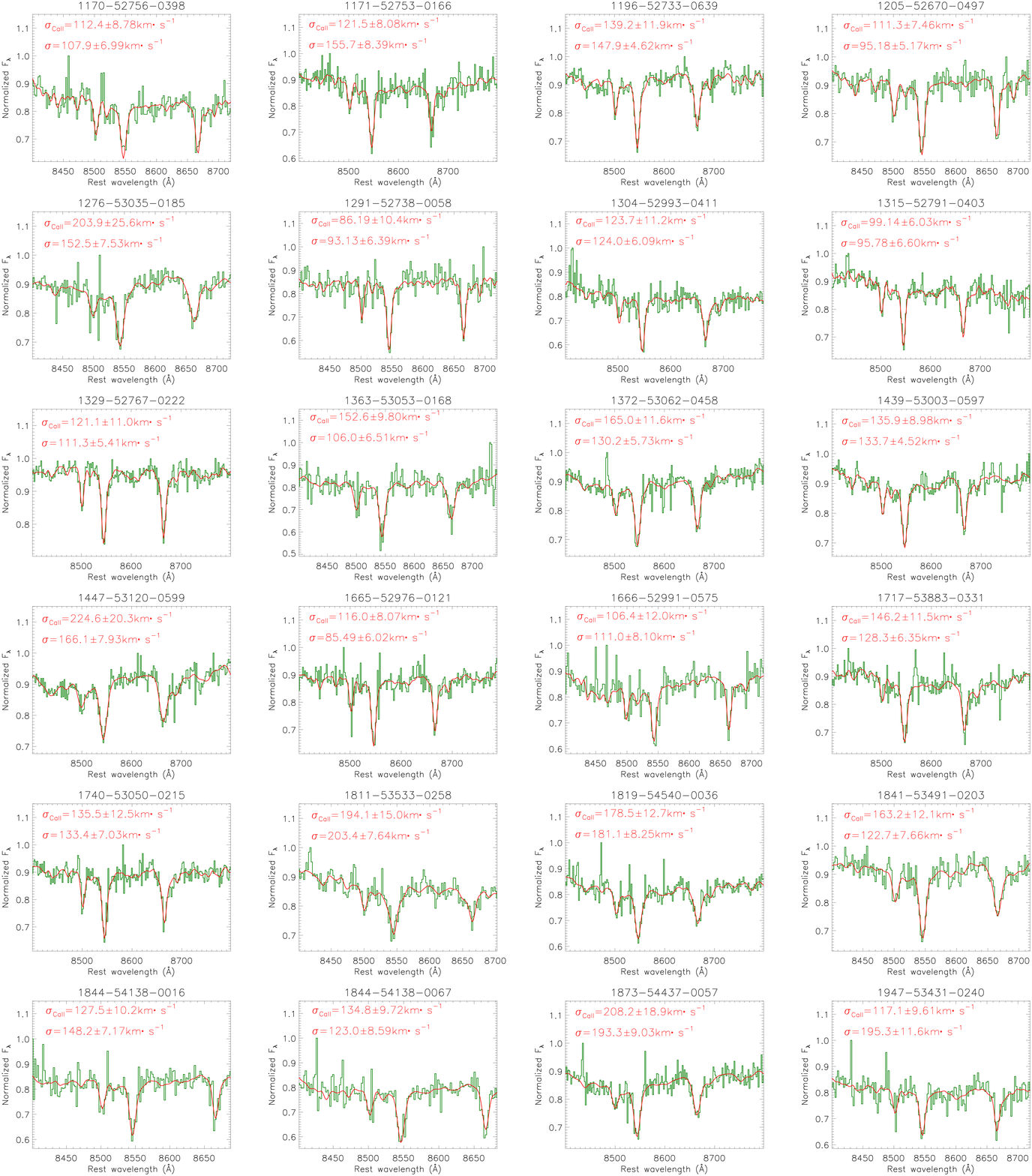}
\caption{--to be continued}
\end{figure*}

\setcounter{figure}{7}
\begin{figure*}
\centering\includegraphics[width = 18cm,height=20cm]{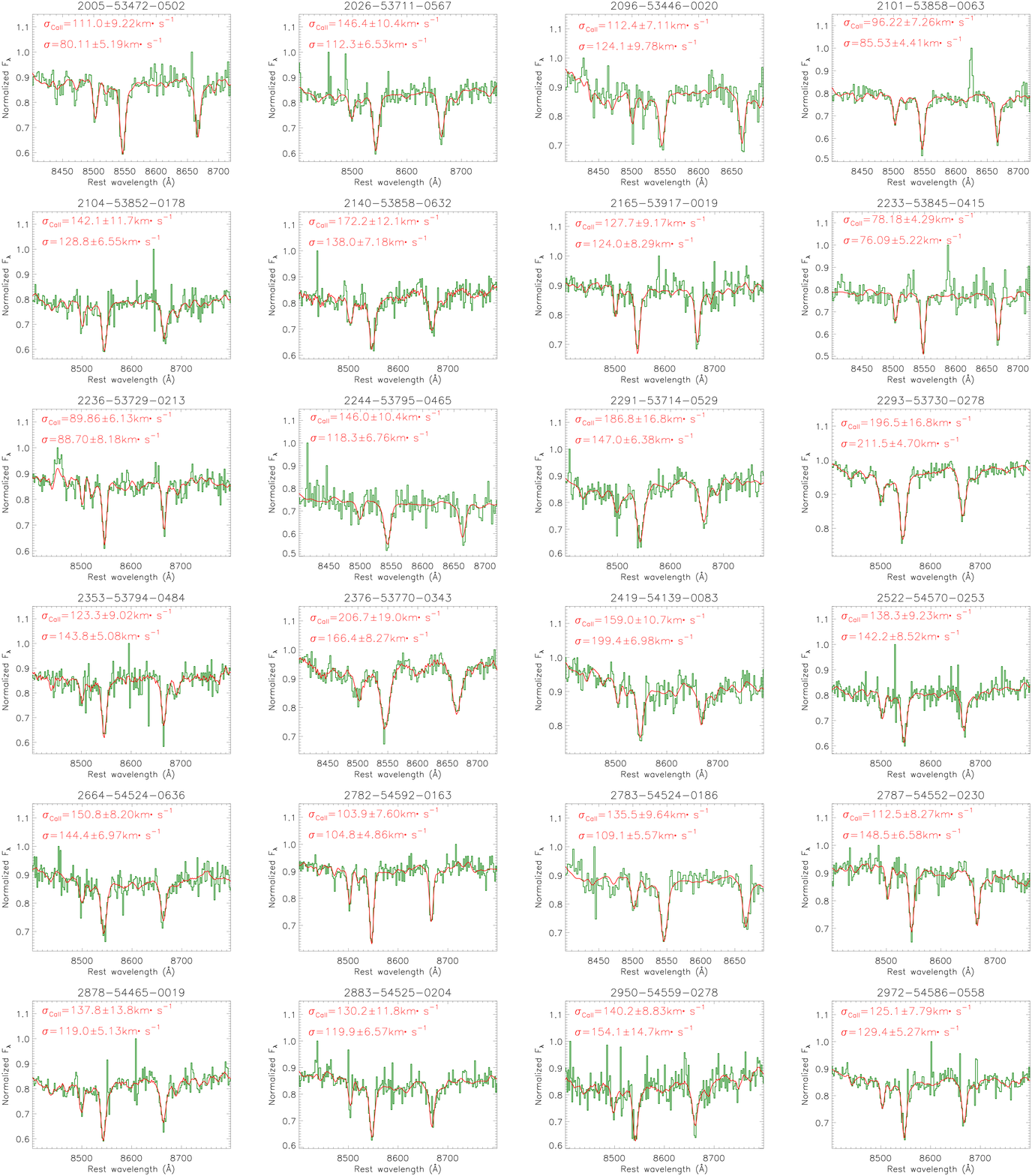}
\caption{--to be continued}
\end{figure*}

\begin{figure} 
\centering\includegraphics[width = 8cm,height=5cm]{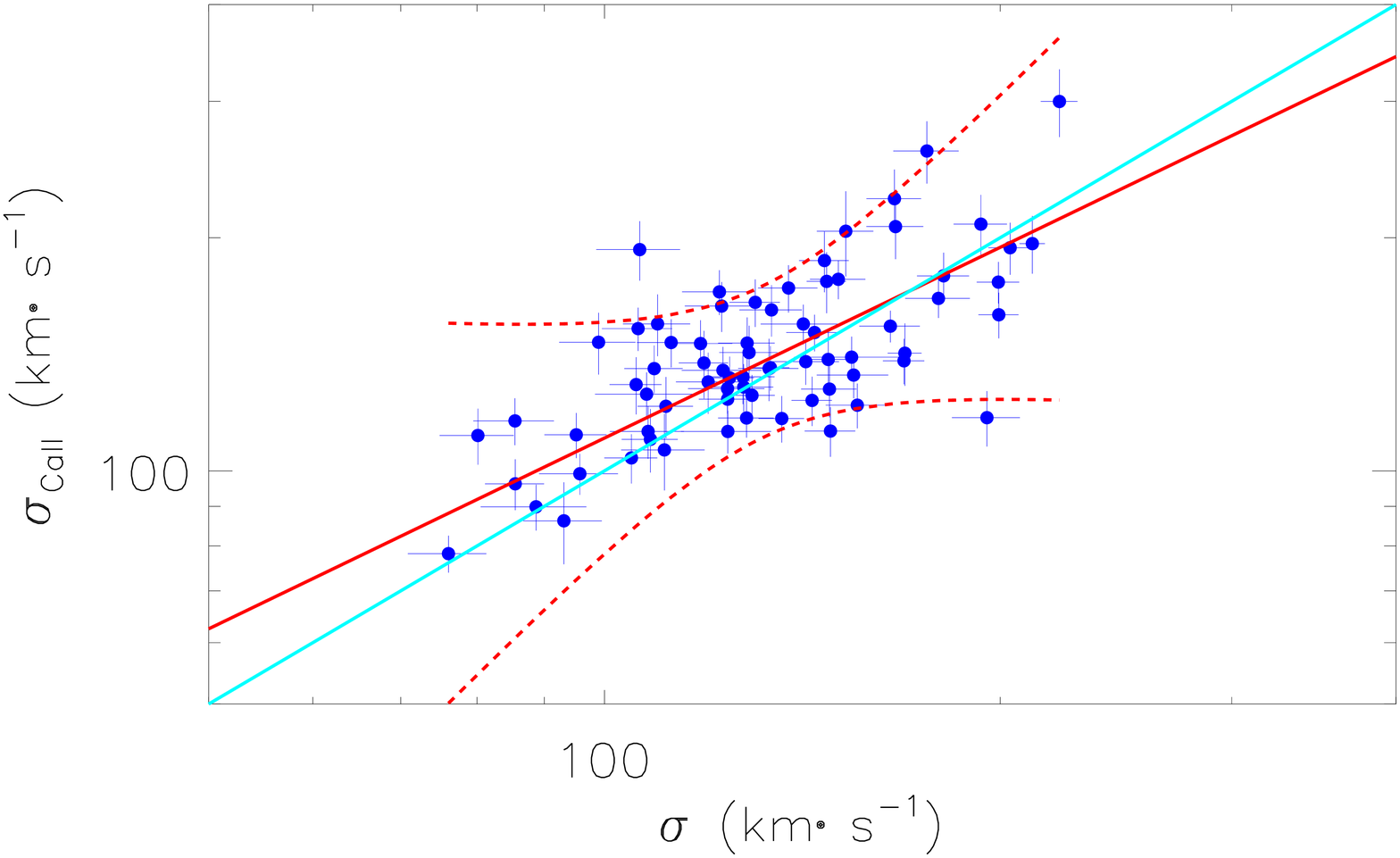}
\caption{Correlation between $\sigma$ and $\sigma_{\rm CaII}$ for the 72 Type-1 AGN with high 
quality Ca~{\sc ii} triplets. Solid red line and dashed red lines show the best fitting results 
and the corresponding 99.99\% confidence bands, and solid blue line shows $\sigma~=~\sigma_{\rm CaII}$.
}
\label{sigma1}
\end{figure}

\begin{figure*}
\centering\includegraphics[width = 8cm,height=5cm]{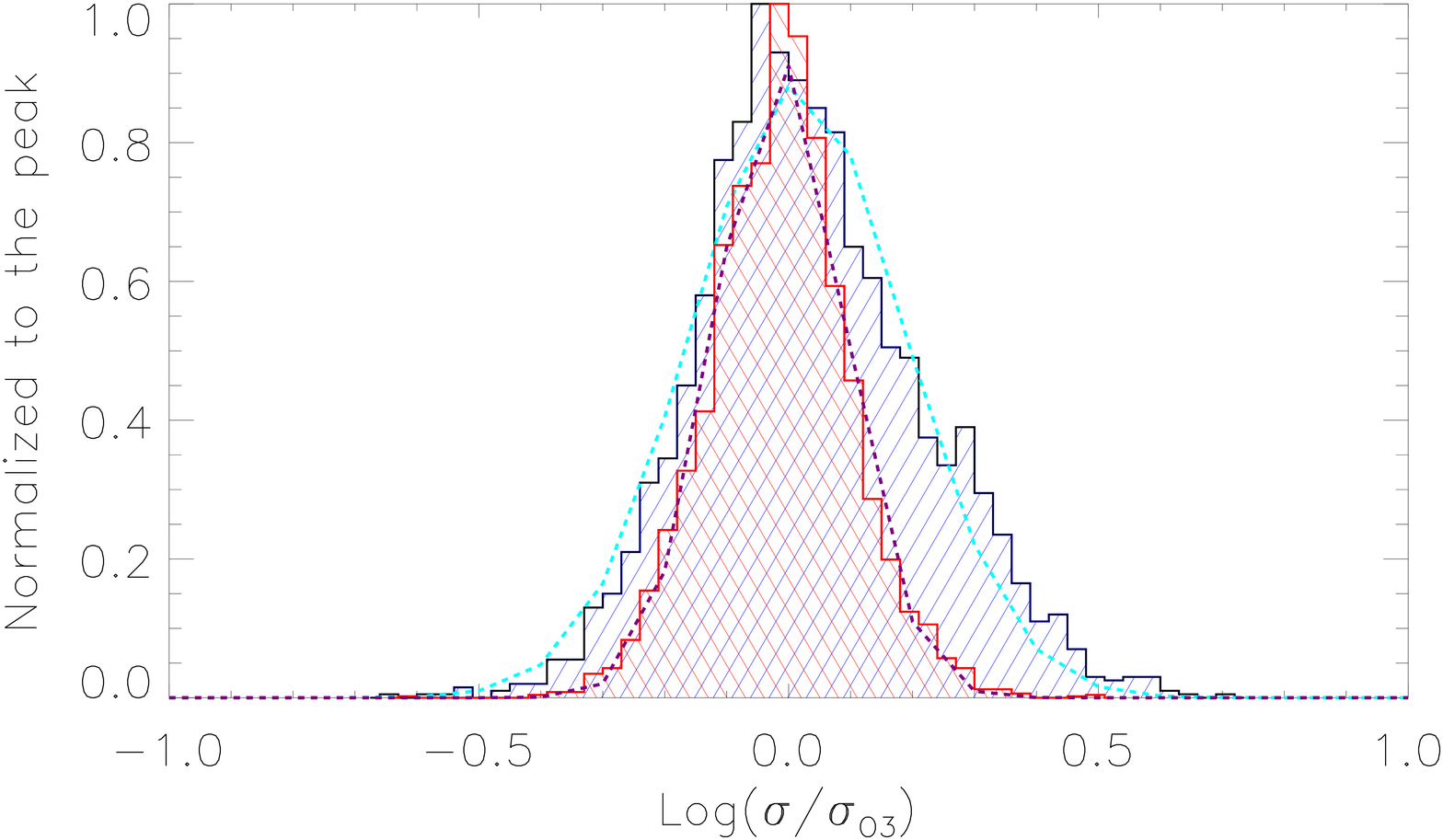}
\centering\includegraphics[width = 8cm,height=5cm]{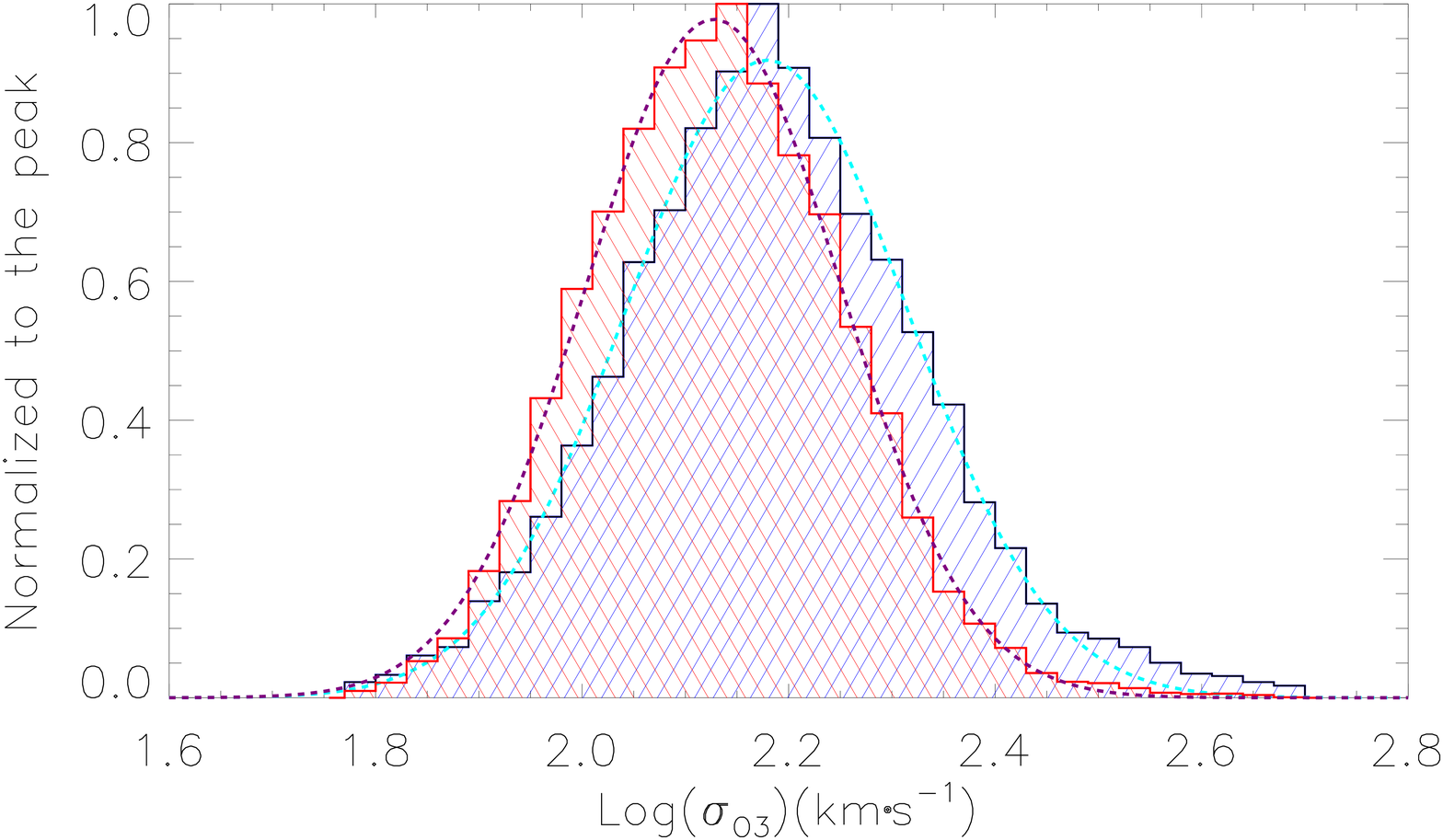}
\centering\includegraphics[width = 8cm,height=5cm]{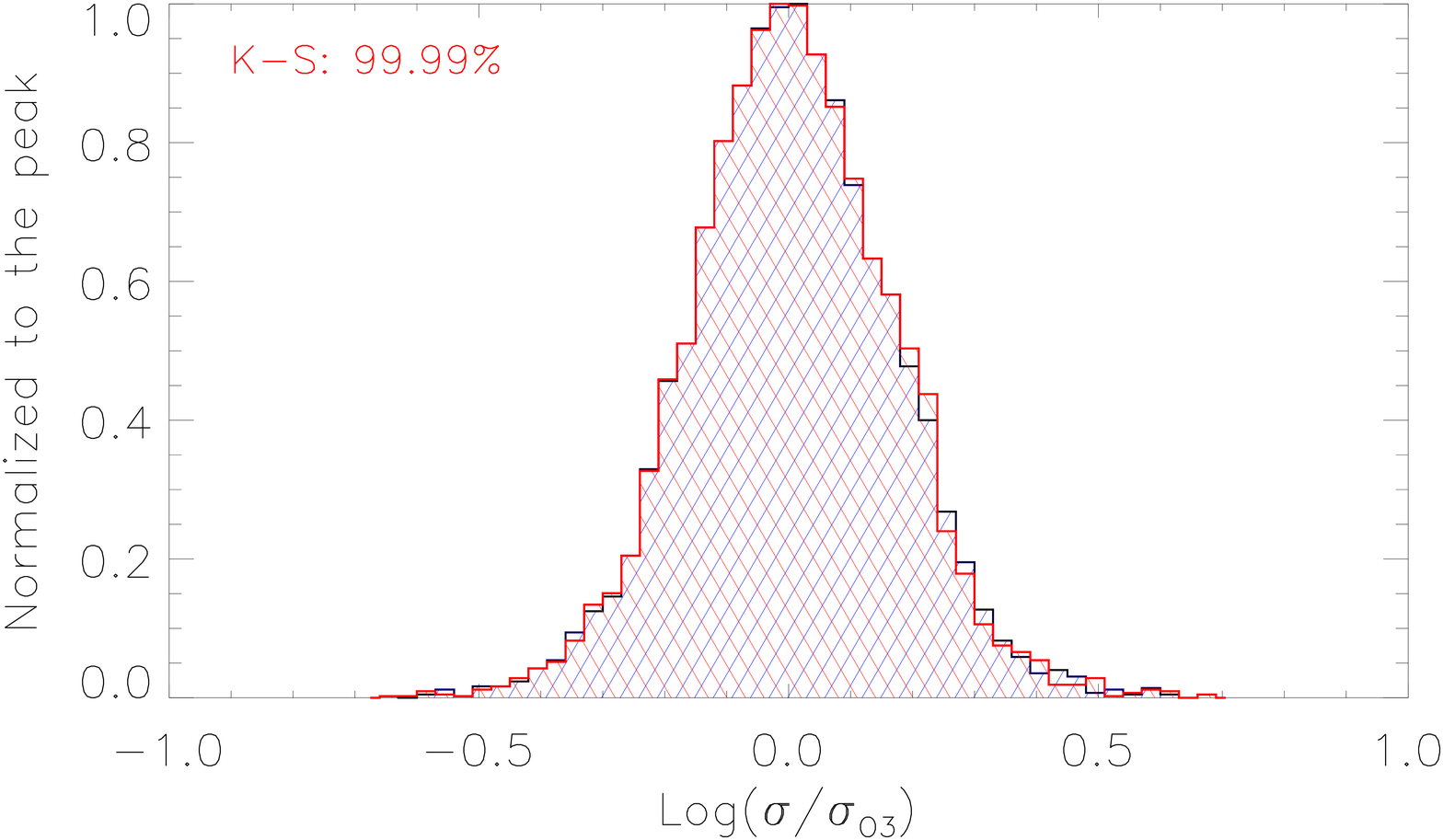}
\centering\includegraphics[width = 8cm,height=5cm]{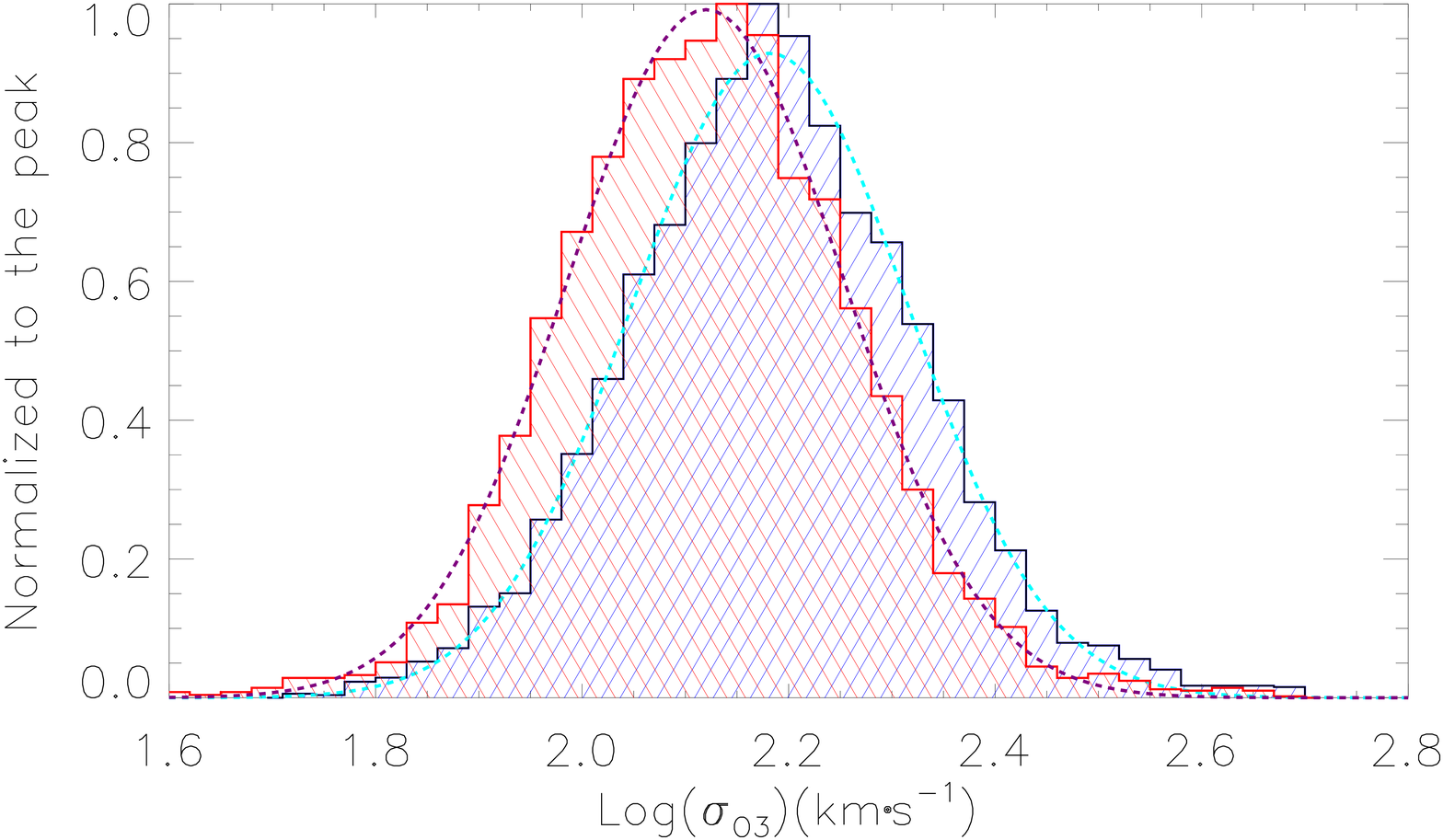}
\caption{Top left panel shows distributions of $\sigma/\sigma_{O3}$ of our measured 
stellar velocity dispersions $\sigma$ to the core [O~{\sc iii}] line width $\sigma_{O3}$ in the 6260 
Type-1 AGN and in the 15353 Type-2 AGN in the main samples. Top right panel shows distributions 
of $\sigma_{O3}$ in the 6260 Type-1 AGN and in the 15353 Type-2 AGN in the main samples. Bottom left 
panel shows distributions of $\sigma/\sigma_{O3}$ in the 5468 Type-1 AGN and in the 5468 Type-2 AGN 
in the subsamples which have the same distributions of $\sigma/\sigma_{O3}$. Bottom right panel  
shows distributions of $\sigma_{O3}$ in the 5468 Type-1 AGN and in the 5468 Type-2 AGN in the 
subsamples which have the same distributions of $\sigma/\sigma_{O3}$. In each panel, histogram filled 
by blue lines and by red lines show distributions of the Type-1 AGN and the Type-2 AGN, respectively, 
dashed line in purple and in cyan represent the corresponding best fitting Gaussian profiles for 
the distributions of the Type-2 AGN and of the Type-1 AGN, respectively. In bottom left panel, the 
Kolmogorov-Smirnov statistic technique provided significance level is marked in red characters.}
\label{so3}
\end{figure*}

	The third comparisons between stellar velocity dispersion and narrow line width are 
applied as indirect evidence to support reliability of the measured stellar velocity dispersions 
in the manuscript. \citet{nw96} have reported a moderately strong correlation between stellar 
velocity dispersion and [O~{\sc iii}] profile width, strongly indicating gravitational motion 
playing an important role in NLRs (narrow emission line regions) velocity field, through a large 
sample of Seyfert galaxies. \citet{ne00} has shown a strong correlation between BH mass and 
[O~{\sc iii}] line width, indicating [O~{\sc iii}] line width can well treated as substitute of 
stellar velocity dispersion. \citet{gh05} have shown that line width of core components of 
[O~{\sc iii}] doublets can trace stellar velocity dispersions, through a large and homogeneous 
sample of SDSS narrow line AGN. \citet{kx07} have shown that the [O~{\sc iii}] line width is a 
good surrogate for stellar velocity dispersions, especially after removal of asymmetric blue wings 
and after excluding [O~{\sc iii}] lines with strong blue-shifts. More recently, \citet{wo16} 
have reported a broad correlation between [O~{\sc iii}] line width and stellar velocity dispersion 
in a large sample of Type-2 AGN, re-confirming that bulge gravitational potential plays a main role 
in NLRs kinematics. \citet{bl18} have shown average ratio of core [O~{\sc iii}] line width to 
stellar velocity dispersion is about 1, but with individual data points off by up to a factor of 
two, through a sample of about 80 SDSS Seyfert 1 galaxies. Therefore, comparisons between our 
measured stellar velocity dispersion $\sigma$ and core [O~{\sc iii}] line width $\sigma_{O3}$ 
are applied, and shown in top left panel of Fig.~\ref{so3}. The mean ratios of 
$\log(\sigma/\sigma_{O3})$ are about 0.035$\pm$0.005 and -0.014$\pm$0.003 for the Type-1 AGN and 
the Type-2 AGN respectively, with uncertainties simply estimated by the bootstrap method with 
1000 loops. The average ratios $\log(\sigma/\sigma_{O3})~\sim~0$ are well consistent with reported 
results in the literature, to provide further clues to support the reliability of the measured 
stellar velocity dispersions in both the Type-2 AGN and the Type-1 AGN in the manuscript.


	Besides the similar mean ratios of $\sigma/\sigma_{O3}\sim1$ between Type-1 AGN and Type-2 
AGN shown in top left panel of Fig.~\ref{so3}, there are different distributions of $\sigma/\sigma_{O3}$ 
between Type-1 AGN and Type-2 AGN, especially the right hand side of the distributions. It is necessary 
and interesting to discuss whether over-estimated stellar velocity dispersions in Type-1 AGN can be 
applied to explain the different distributions. As is well known, there are few  
contaminations of stellar absorption features or continuum emissions or optical broad Fe~{\sc ii} 
emissions on measured line width $\sigma_{O3}$ of core components of [O~{\sc iii}] emission lines 
both in Type-1 AGN as shown in \citet{sh11} and in Type-2 AGN as shown in \citet{gh05}. Top right panel 
of Fig.~\ref{so3} shows $\sigma_{O3}$ distributions of Type-1 AGN and Type-2 AGN. There are higher 
$\sigma_{O3}$ in Type-1 AGN with mean $\log(\sigma_{O3}/{\rm km\cdot~s^{-1}})$ about 2.179$\pm$0.005 
(about $151\pm2{\rm km\cdot~s^{-1}}$) than in Type-2 AGN with mean $\log(\sigma_{O3}/{\rm km\cdot~s^{-1}})$ 
about 2.127$\pm$0.004 (about $133\pm2{\rm km\cdot~s^{-1}}$). Besides the different $\sigma_{O3}$ 
distributions shown in top right panel of Fig.~\ref{so3} for the samples of Type-1 AGN and Type-2 
AGN with quite different distributions of $\sigma/\sigma_{O3}$, two subsamples\footnote{It is quite 
easy to create the two subsamples, based on the shown distributions of $\sigma/\sigma_{O3}$ in top 
left panel of Fig.~\ref{so3}. Here, there are no detailed descriptions on how to create the subsamples, 
but quite similar as what will be done in Section 5.3.} of 5468 Type-1 AGN and 5468 Type-2 AGN 
are collected to have the same distributions of $\sigma/\sigma_{O3}$ which is shown in bottom left 
panel of Fig.~\ref{so3} with significance levels higher than 99.99\% through the two-sided 
Kolmogorov-Smirnov statistic technique. If there were over-estimated stellar velocity dispersions 
in Type-1 AGN leading to the different distributions shown in top left panel of Fig.~\ref{so3}, 
there should be quite smaller $\sigma_{O3}$ difference between Type-1 AGN and Type-2 AGN in the 
subsamples than in the main samples. However, as shown distributions in bottom right panel of 
Fig.~\ref{so3} of $\sigma_{O3}$ in the Type-1 AGN and Type-2 AGN in the subsamples which have the 
same distributions of $\sigma/\sigma_{O3}$, the mean values $\log(\sigma_{O3}/{\rm km\cdot~s^{-1}})$ 
are about 2.182$\pm$0.005 and 2.119$\pm$0.004\ in the Type-1 AGN and in the Type-2 AGN in the 
subsamples, respectively. The quite similar mean values of $\sigma_{O3}$ between Type-1 AGN and 
Type-2 AGN in the subsamples and in the main samples strongly indicate that the different distributions 
of $\sigma/\sigma_{O3}$ shown in top left panel of Fig.~\ref{so3} are not due to over-estimated stellar 
velocity dispersions in Type-1 AGN, but mainly due to different distributions of $\sigma_{O3}$ between 
Type-1 AGN and Type-2 AGN. Therefore, results in Fig.~\ref{so3} can be accepted as indirect evidence 
to support the reliability of our measured stellar velocity dispersions, but there are no further 
discussions on the different distributions of $\sigma_{O3}$ in Type-1 AGN and Type-2 AGN which 
is beyond the scope of the manuscript and have few effects on our final conclusions in the manuscript.

\begin{figure*} 
\centering\includegraphics[width = 18cm,height=6cm]{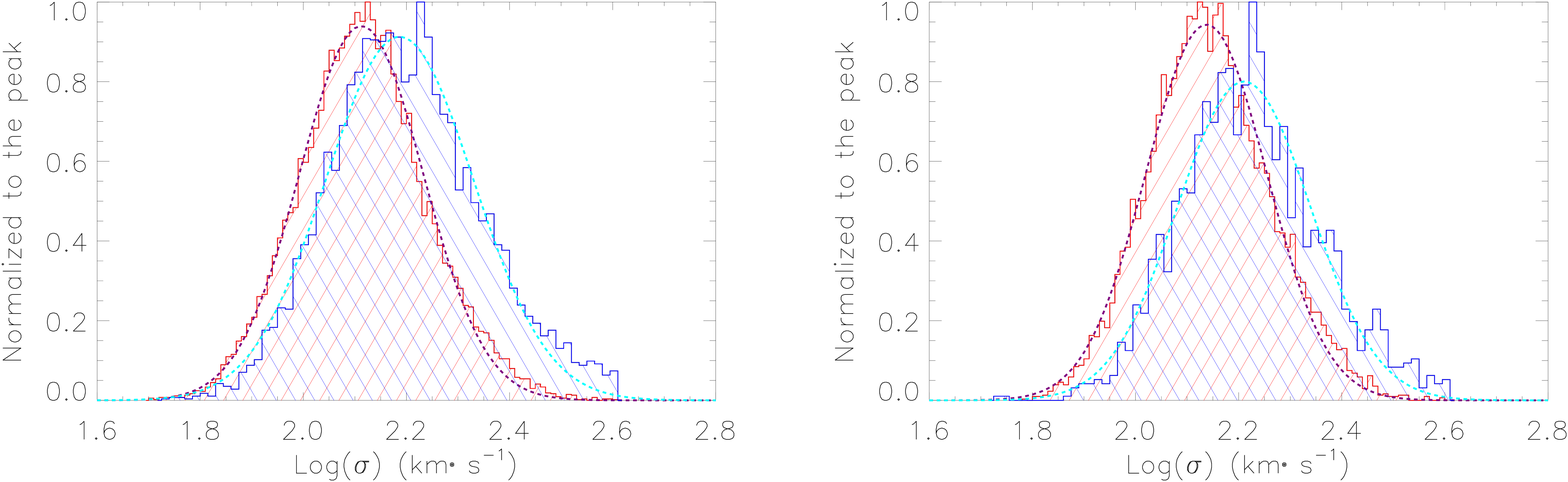}
\caption{Left panel shows distributions of stellar velocity dispersions of the 6260 Type-1 
AGN (histogram filled by blue lines) and the 15353 Type-2 AGN (histogram filled by red lines), 
respectively. Right panel shows distributions of high quality stellar velocity dispersions 
(10 times larger than their uncertainties) of the 1680 Type-1 AGN (histogram filled by blue 
lines) and the 8535 Type-2 AGN (histogram filled by red lines), respectively. In each panel, 
dashed lines in purple and in cyan represent the corresponding best fitting Gaussian profiles 
for the $\log(\sigma)$ distributions of the Type-2 AGN and of the Type-1 AGN, respectively.}
\label{sigma12}
\end{figure*}

\section{Main Results and Discussions}

\subsection{Direct comparisons of stellar velocity dispersions}

	Based on the measured $\sigma$ through absorption features around 4000\AA, left panel 
of Fig.~\ref{sigma12} shows the direct $\sigma$ comparisons between the largest sample of 6260 
Type-1 AGN and the largest sample of 15353 Type-2 AGN in SDSS DR12. The mean $\log(\sigma)$ are 
about $\log(\sigma/{\rm km\cdot~s^{-1}})~\sim~2.199\pm0.006$ ($\sigma~\sim~158\pm3 {\rm km\cdot~s^{-1}}$) 
and $\log(\sigma/{\rm km\cdot~s^{-1}})~\sim~2.117\pm$0.004 ($\sigma~\sim~132\pm2 {\rm km\cdot~s^{-1}}$) 
for the Type-1 AGN and the Type-2 AGN, respectively, indicating statistically larger stellar velocity 
dispersions in Type-1 AGN. Uncertainty of each mean value is determined by the bootstrap method with 
1000 loops. Meanwhile, in order to ensure few effects of data quality on the different distributions 
of $\log(\sigma)$, there are 1680 Type-1 AGN and 8535 Type-2 AGN collected through the criterion that 
the measured stellar velocity dispersions at least 10 times larger than their uncertainties. The 
corresponding comparison results on the high quality stellar velocity dispersions are shown in right 
panel of Fig.~\ref{sigma12}, with mean $\log(\sigma)$ about 
$\log(\sigma/{\rm km\cdot~s^{-1}})\sim2.222\pm0.003$ ($\sigma~\sim~167\pm3 {\rm km\cdot~s^{-1}}$) 
and $\log(\sigma/{\rm km\cdot~s^{-1}})\sim2.144\pm0.002$ ($\sigma~\sim~138\pm1 {\rm km\cdot~s^{-1}}$) 
for the Type-1 AGN and the Type-2 AGN, respectively, with uncertainty of each mean value determined 
by the bootstrap method with 1000 loops.

	Moreover, considering the discussed results in Section 3 that the measured stellar velocity 
dispersions in Type-1 AGN are about 4\% larger than the intrinsic values, the corrected mean 
$\sigma$ are about $\sigma~\sim~152\pm2 {\rm km\cdot~s^{-1}}$ and 
$\sigma~\sim~161\pm2 {\rm km\cdot~s^{-1}}$ for the 6260 Type-1 AGN and for the 1680 Type-1 AGN with 
high quality stellar velocity dispersions, respectively, to re-confirm statistically larger stellar 
velocity dispersions in Type-1 AGN. After considering the measured stellar velocity dispersions in 
Type-1 AGN about 4\% larger than the intrinsic values, the Students T-statistic technique is re-applied 
to determine the different mean values of $\log(\sigma)$ between Type-1 AGN (with measured stellar 
velocity dispersions scaled by 0.96) and Type-2 AGN with confidence level higher than 10sigma, and 
the two-sided Kolmogorov-Smirnov statistic technique indicates stellar velocity dispersions of the 
Type-1 AGN and the Type-2 AGN obey the same distributions of $\log(\sigma)$ with significance level 
smaller than $10^{-25}$. Therefore, in spite of the following necessary discussed effects, the basic 
results can be found that Type-1 AGN have their stellar velocity dispersions statistically 16\% larger 
than Type-2 AGN.

\begin{figure}
\centering\includegraphics[width = 8cm,height=5cm]{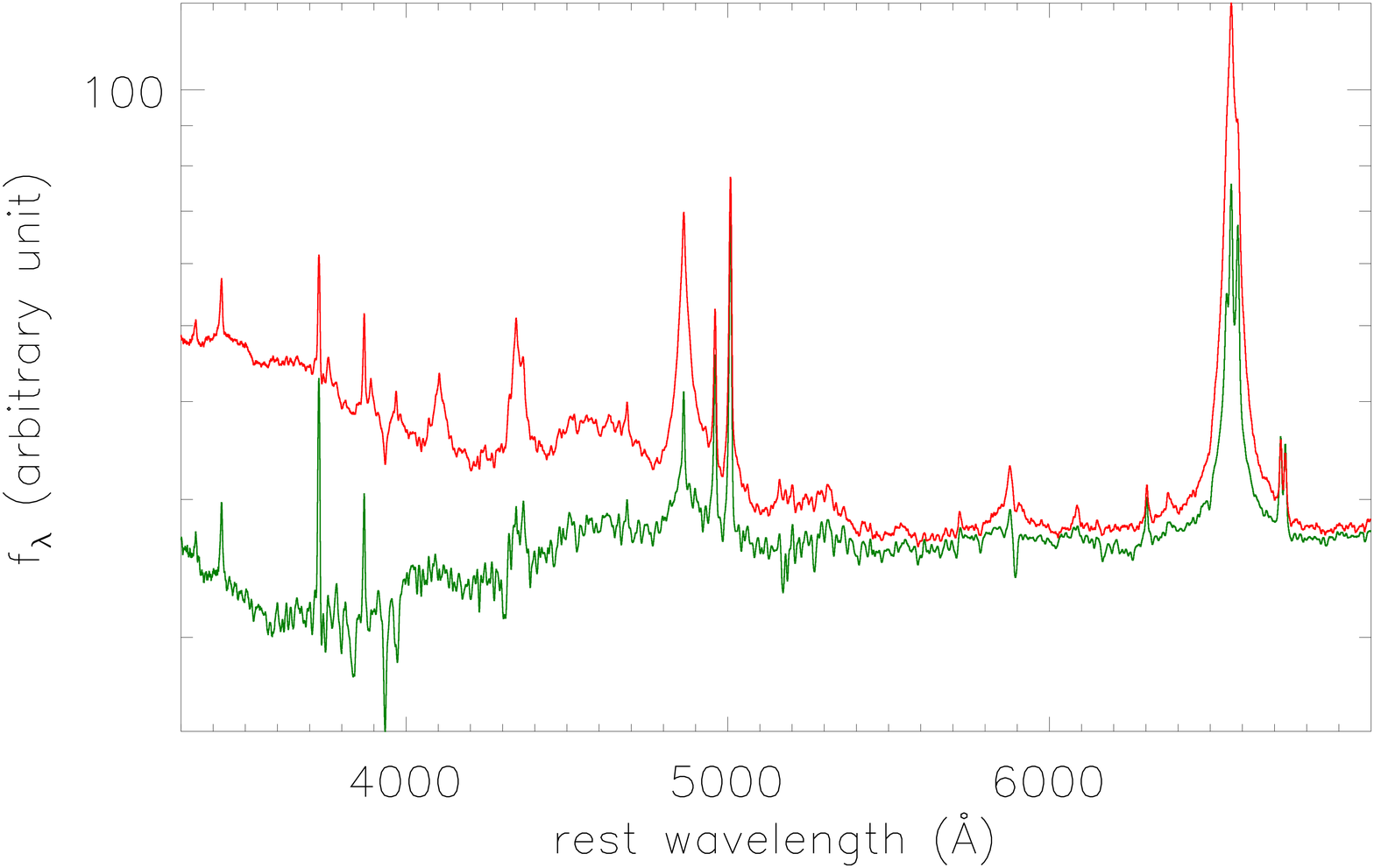}
\caption{The inverse variance weighted mean spectra of Type-1 AGN (solid dark green line) with 
stellar velocity dispersions well measured through absorption features around 4000\AA~ and of the 
other Type-1 AGN (solid red line) without apparent absorption features around 4000\AA.
}
\label{msp}
\end{figure}

\begin{figure*}  
\centering\includegraphics[width = 18cm,height=6cm]{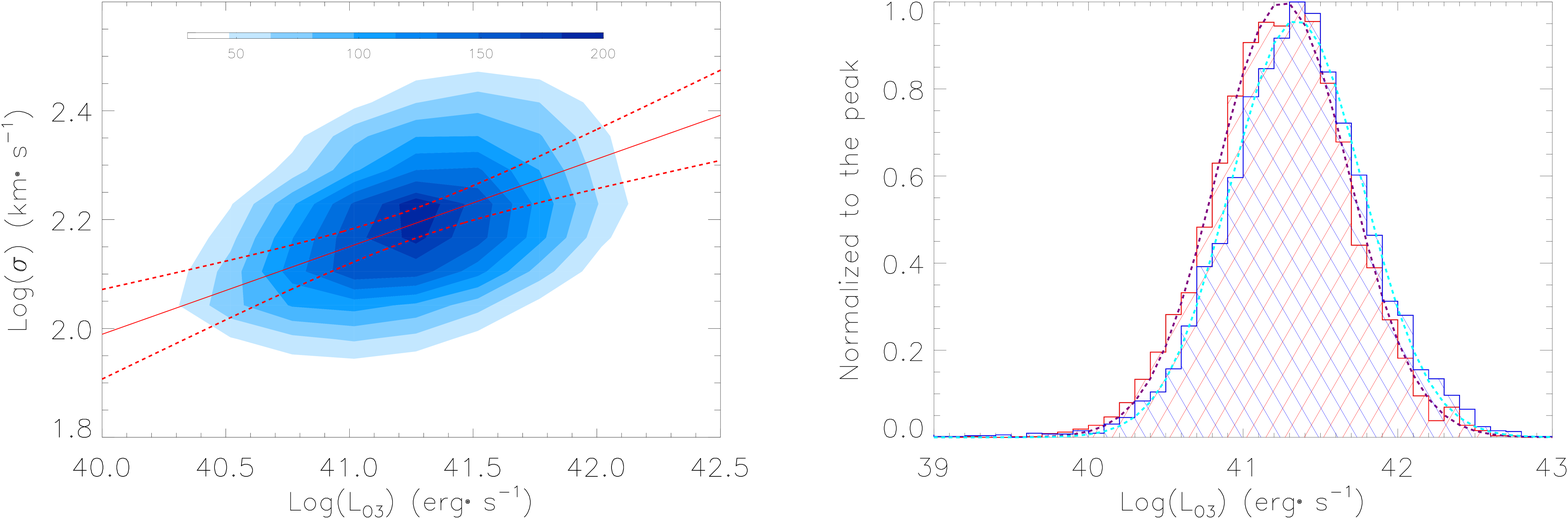}
\caption{Left panel shows dependence of $\sigma$ on total [O~{\sc iii}] line luminosity $L_{O3}$ 
of the 6260 Type-1 AGN with well measured $\sigma$. Solid red line and dashed red lines show the 
best-fitting results and the corresponding 5sigma confidence bands, respectively. Right panel shows 
$L_{O3}$ distributions of the Type-1 AGN with (histogram filled with red lines) and without 
(histogram filled with blue lines) apparent absorption features around 4000\AA. Dashed lines in 
purple and in cyan represent the corresponding best-fitting Gaussian profiles for the $L_{O3}$ 
distributions of the Type-1 AGN with and without measured stellar velocity dispersions, respectively.
}
\label{sL3}
\end{figure*}

\begin{figure*} 
\centering\includegraphics[width = 18cm,height=6cm]{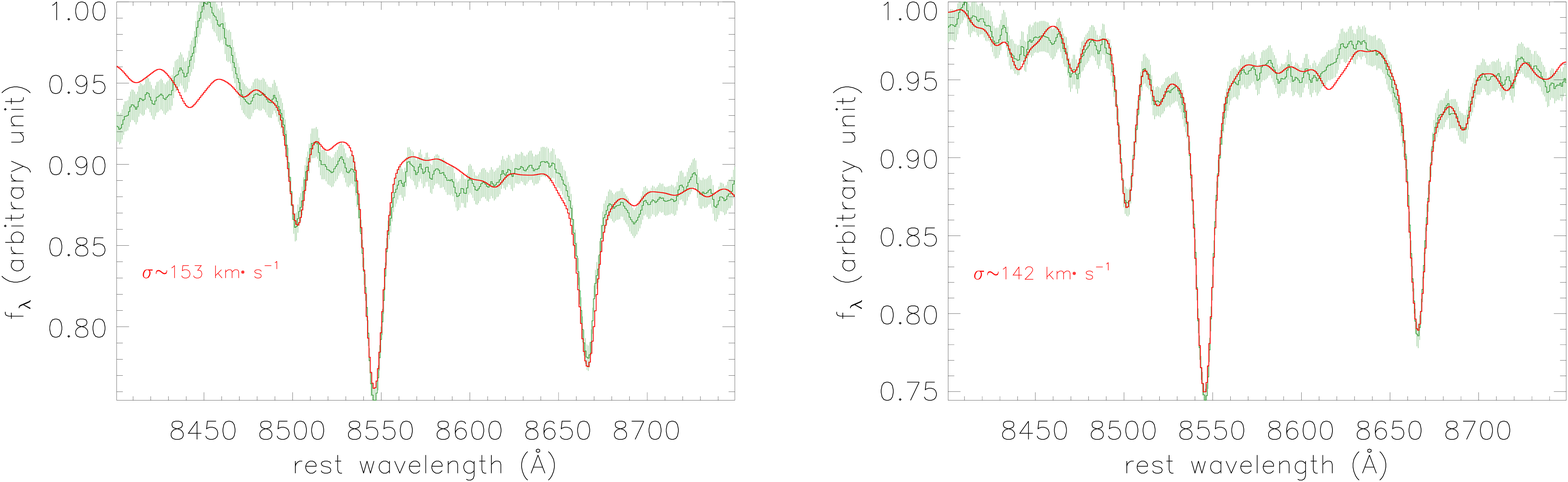}
\caption{The mean spectra around 8500\AA~ of the Type-1 AGN without (left panel) and with (right 
panel) apparent absorption features around 4000\AA. Solid red line shows the best-fitting results 
to Ca~{\sc ii} triplet. The measured stellar velocity dispersion through Ca~{\sc ii} triplet is 
marked in red characters in each panel.	
}
\label{msca}
\end{figure*}

\begin{figure*} 
\centering\includegraphics[width = 18cm,height=6cm]{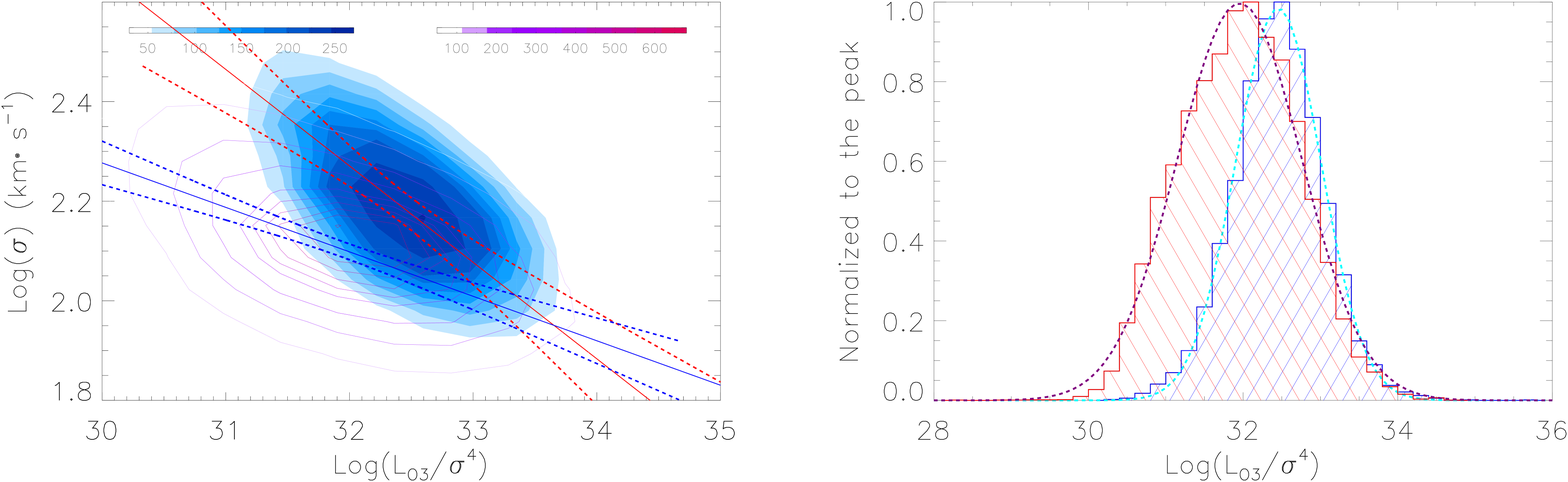}
\caption{Left panel shows dependence of $\sigma$ on $L_{O3}/\sigma^4$ for the Type-1 AGN 
(contour filled by bluish colors) and the Type-2 AGN (contour levels by reddish lines). Solid 
red line and dashed red lines are for the best-fitting results and the corresponding 5sigma 
confidence bands for the Type-1 AGN. Solid blue line and dashed blue lines are for the 
best-fitting results and the corresponding 5sigma confidence bands for the Type-2 AGN. Right 
panel shows $L_{O3}/\sigma^4$ distributions of the Type-1 AGN (histogram filled with blue 
lines) and Type-2 AGN (histogram filled with red lines). 
}
\label{acc}
\end{figure*}

\subsection{Dependence of stellar velocity dispersion on [O~{\sc iii}] luminosity}

	As the shown results in section 2, there are half of Type-1 AGN of which stellar velocity 
dispersions $\sigma$ can not be well measured. Therefore, in the subsection, necessary discussions 
are given on expected probable properties of the intrinsic stellar velocity dispersions of the 
half of Type-1 AGN without measured $\sigma$ in the parent sample.

	There is only one probability leading to stellar velocity dispersions not be well measured 
through absorption features around 4000\AA~ in Type-1 AGN, the host galaxy contributions are 
weak enough that absorption features around 4000\AA~ are overwhelmed in emission features of 
AGN activities. Therefore, it is interesting to check spectral features of the Type-1 AGN with 
and without measured $\sigma$. Fig.~\ref{msp} shows the inverse variance weighted mean spectra of 
the 6260 Type-1 AGN with well measured $\sigma$ and of the other Type-1 AGN with $\sigma$ not 
measured due to unapparent absorption features around 4000\AA. The mean spectra with bluer 
continuum emissions around 4000\AA~ can be well applied to confirm that the half of Type-1 AGN 
without measured $\sigma$ are mainly due to unapparent absorption features around 4000\AA~ which 
are overwhelmed by emission features of central AGN activities.

	Meanwhile, left panel Fig.~\ref{sL3} shows dependence of measured $\sigma$ on total 
[O~{\sc iii}] line luminosity $L_{O3}$ of the 6260 Type-1 AGN with well measured $\sigma$. 
The dependence could provide clues to expected properties of intrinsic $\sigma$ of the other 
half of Type-1 AGN without measured $\sigma$ in the parent sample. There is a positive linear 
correlation between $\sigma$ and $L_{O3}$ with Spearman Rank correlation coefficient of 0.38 
with $P_{null}~<~10^{-10}$. And the positive linear correlation can be described by 
\begin{equation}
\log(\frac{\sigma}{\rm km\cdot~s^{-1}})~\sim~-4.45~+~
	0.16\times\log(\frac{L_{O3}}{\rm erg\cdot~s^{-1}})
\end{equation}
indicating that Type-1 AGN with higher [O~{\sc iii}] line luminosity will have statistically 
larger intrinsic $\sigma$. Then, right panel of Fig.~\ref{sL3} shows distributions of 
$\log(L_{O3}/{\rm erg\cdot~s^{-1}})$ with mean values of 
$~\sim~41.245\pm0.006$ ($L_{O3}\sim(1.76\pm0.02)\times10^{41}{\rm erg\cdot~s^{-1}}$) and 
$~\sim~41.340\pm0.008$ ($L_{O3}\sim(2.19\pm0.04)\times10^{41}{\rm erg\cdot~s^{-1}}$) of the 
Type-1 AGN with and without well measured $\sigma$. The uncertainties of the mean values are 
estimated by the bootstrap method with 1000 loops. And the two-sided Kolmogorov-Smirnov statistic 
technique indicates the two distributions of $\log(L_{O3})$ obey the same distributions with 
significance level only about $10^{-22}$. Therefore, the half of Type-1 AGN without measured 
$\sigma$ should have statistically larger $\sigma$ (at least not smaller) than the 6260 Type-1 
AGN with apparent absorption features around 4000\AA. Therefore, considering the half of Type-1 
AGN without well measured $\sigma$, the mean $\sigma$ of all the 12342 Type-1 AGN in the parent 
sample should be larger than the ones shown in Fig.~\ref{sigma12}.

	Moreover, although only dozens of Type-1 AGN have Ca~{\sc ii} triplets in SDSS spectra, 
mean spectra around 8500\AA~ of the low redshift Type-1 AGN with and without measured $\sigma$ 
are well checked, in order to examine whether the Type-1 AGN without measured $\sigma$ through 
absorption features around 4000\AA~ have statistically larger $\sigma$, based on the $\sigma$ 
measured through Ca~{\sc ii} triplets. Fig.~\ref{msca} shows the mean spectra around 8500\AA~ 
of Type-1 AGN with and without measured $\sigma$ through absorption features around 4000\AA, 
and the best fitting results by single stellar template as what have been done to absorption 
features around 4000\AA. The determined stellar velocity dispersions are about 
$\sigma_{CaII}~\sim~153{\rm km\cdot~s^{-1}}$ and $\sigma_{CaII}~\sim~142{\rm km\cot~s^{-1}}$ 
for the Type-1 AGN without and with apparent absorption features around 4000\AA, respectively. 
The results are consistent with the expected larger $\sigma$ of the Type-1 AGN without apparent 
absorption features around 4000\AA. Therefore, considering the half Type-1 AGN without measured 
$\sigma$ will lead to statistically larger $\sigma$ for all the 12342 Type-1 AGN than those shown 
in Fig.~\ref{sigma12} for the 6260 Type-1 AGN in the main sample.

	Furthermore, as discussed in \citet{kb06}, ratio of [O~{\sc iii}] line luminosity to 
$\sigma^4$ (applied to trace central BH mass) can be accepted as a good indicator of central 
accretion rate relative to the Eddington ratio, leading to continuous sequence for different 
subclasses of Type-2 AGN. Here, properties of $L_{O3}/\sigma^4$ are checked for Type-1 AGN and 
Type-2 AGN, with $L_{O3}$ in unit of ${\rm erg\cdot~s^{-1}}$ and $\sigma$ in unit of 
${\rm km\cdot~s^{-1}}$.

	Before proceeding further, as discussed results on extended components of [O~{\sc iii}] 
doublets in AGN in \citet{zh17, zh21m}, not line luminosity of total [O~{\sc iii}]$\lambda5007$\AA~ 
but of the core component of [O~{\sc iii}]$\lambda5007$\AA~ is applied to check the properties 
of $L_{O3}/\sigma^4$ of Type-1 AGN and Type-2 AGN, because of seriously obscured extended 
components of [O~{\sc iii}]$\lambda5007$\AA~ in Type-2 AGN and of the strong linear correlation 
between AGN continuum luminosity and the luminosity of the core component of 
[O~{\sc iii}]$\lambda5007$\AA. Properties of $L_{O3}/\sigma^4$ are shown in Fig.~\ref{acc}. 
Different $L_{O3}/\sigma^4$ can be confirmed between Type-1 AGN and Type-2 AGN. There are positive 
linear correlations between $L_{O3}/\sigma^4$ and $\sigma$ with Spearman rank correlation 
coefficients of about 0.38 and 0.31 with $P_{null}~<~10^{-10}$ for the Type-1 AGN and the Type-2 AGN, 
and the linear correlations can be described by 
\begin{equation}
\begin{split}
\log(\frac{\sigma}{\rm km\cdot~s^{-1}}) &\sim~8.47
	~-~0.19\times\log(\frac{L_{O3}}{\sigma^4})\ \ (Type-1) \\
\log(\frac{\sigma}{\rm km\cdot~s^{-1}}) &\sim~4.96
	~-~0.09\times\log(\frac{L_{O3}}{\sigma^4})\ \ (Type-2)
\end{split}
\end{equation}
The mean $\log(L_{O3}/\sigma^4)$ are about 32.448$\pm$0.009 and 31.948$\pm$0.008 of the Type-1 
AGN and the Type-2 AGN respectively, with the uncertainties estimated by the bootstrap method 
with 1000 loops. Therefore, Type-1 AGN and Type-2 AGN have quite different properties of 
$\log(L_{O3}/\sigma^4)$. Actually, line luminosity of total [O~{\sc iii}]$\lambda5007$\AA~ can 
lead to the similar results on the quite different ratios of total [O~{\sc iii}] line luminosity 
to $\sigma^4$ in Type-1 AGN and Type-2 AGN. Different $\log(L_{O3}/\sigma^4)$ provide different 
central activity properties of the collected Type-1 AGN and Type-2 AGN, indicating further effects 
should be carefully considered on the shown results in Fig.~\ref{sigma12}.

\begin{figure*} 
\centering\includegraphics[width = 18cm,height=21cm]{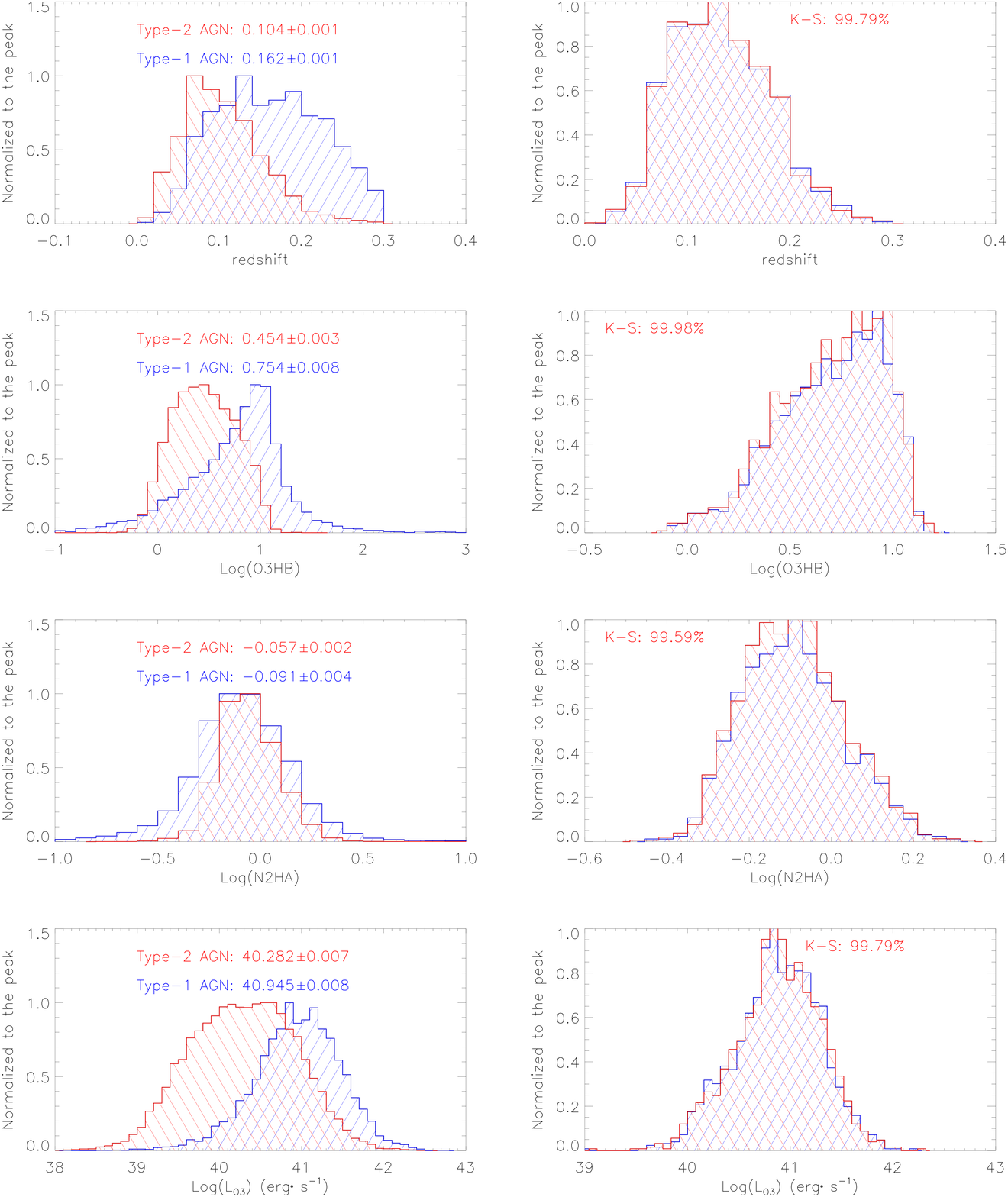}
\caption{Distributions of redshift (top panels), of narrow line flux ratios of O3HB and N2HA 
(middle two panels) and of luminosity of core components of [O~{\sc iii}] line (bottom panels) 
of the Type-1 AGN (histogram filled with blue lines) and Type-2 AGN (histogram filled with red 
lines) in the main samples (left panels) and in the subsamples (right panels). In each right 
panel, the Kolmogorov-Smirnov statistic technique provided significance level is marked in red 
characters. In each left panel, the mean values of the Type-1 AGN and the Type-2 AGN are marked 
in blue and red characters in top region.
}
\label{dis}
\end{figure*}

\begin{figure*}
\centering\includegraphics[width = 18cm,height=7cm]{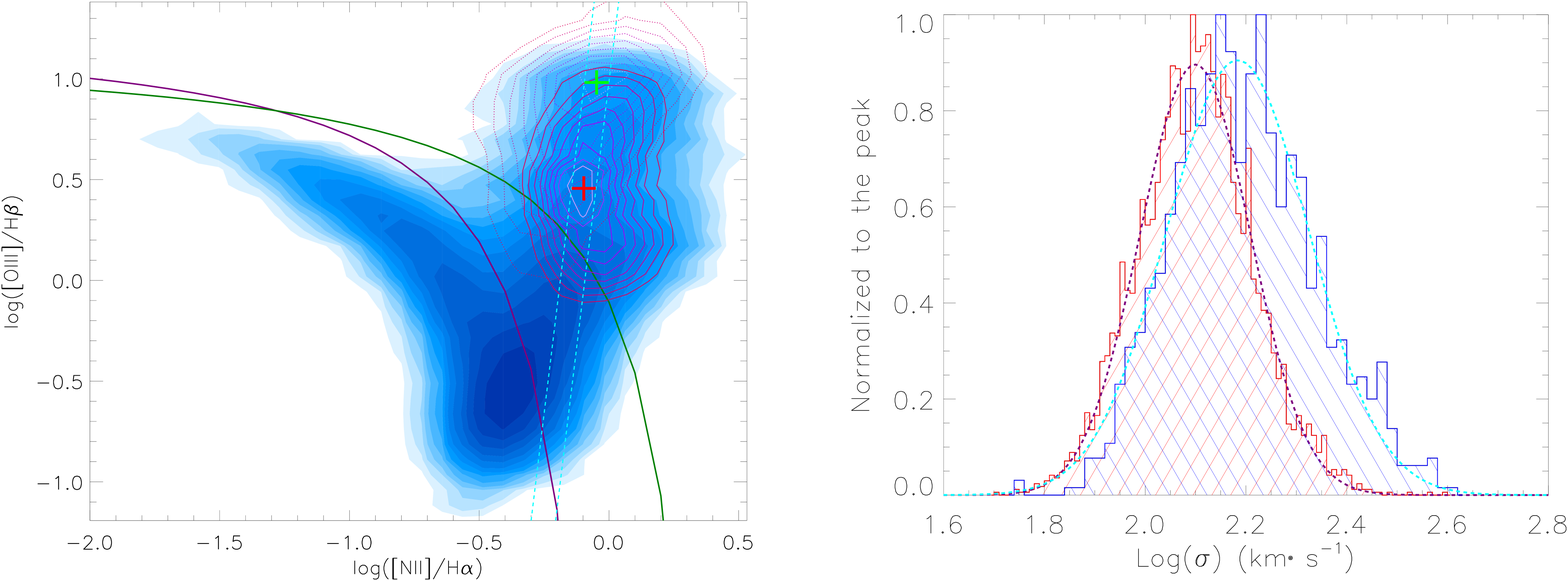}
\caption{Left panel shows type-1 AGN and Type-2 AGN in the BPT diagram of O3HB versus N2HA. 
Contour filled with bluish colors represents the results for all the narrow emission line 
galaxies collected from SDSS DR12. Contour with levels as solid reddish lines and with levels 
as dotted reddish lines represent the results for the Type-2 AGN and the Type-1 AGN with reliable 
stellar velocity dispersions in the manuscript, respectively. Thick pluses in red and in green 
mark the central positions of the contours for the Type-2 AGN and for the Type-1 AGN, respectively. 
Solid purple line and solid dark green line show the dividing lines between different kinds 
of narrow emission line galaxies in \citet{kb06, ka03a}. Dashed cyan lines mark the strip 
applied to collect Type-1 AGN and Type-2 AGN to simply check effects of AGN activities on 
stellar velocity dispersion comparisons. Right panel shows the results Similar as Fig.~\ref{sigma12}, 
but for the 1044 Type-1 AGN and the 4202 Type-2 AGN covered in the strip shown in left panel. 
The symbols and line styles have the same meanings as those in Fig.~\ref{sigma12}.
}
\label{bpt}
\end{figure*}

\begin{figure*}
\centering\includegraphics[width = 15cm,height=10cm]{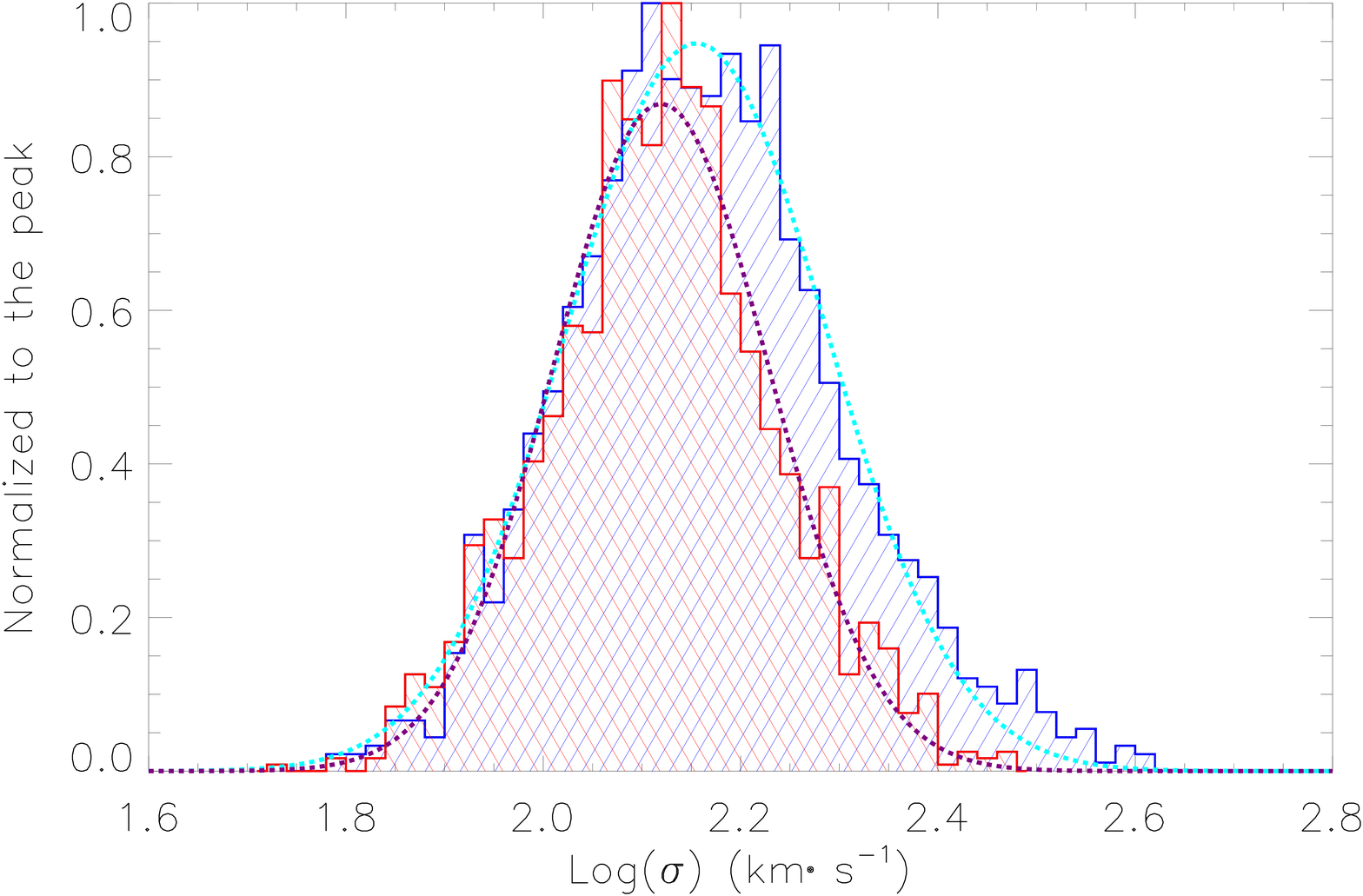}
\caption{Similar as Fig.~\ref{sigma12}, but for the 1445 Type-1 AGN and the 1445 Type-2 AGN in 
the subsamples, which have the same distributions of $z$, O3HB, N2HA and $L_{O3}$. The symbols 
and line styles have the same meanings as those in Fig.~\ref{sigma12}.}
\label{dis2}
\end{figure*}

\subsection{Necessary effects on the comparisons of stellar velocity dispersions}

	Effects of different redshift are firstly considered on the results shown in 
Fig.~\ref{sigma12}, because different redshift provides different evolutionary histories of 
host galaxies and central BHs. Different distributions of redshift of Type-1 AGN and Type-2 
AGN are shown in top left panel of Fig.~\ref{dis}. Mean redshifts are about 0.162$\pm$0.001 
and 0.104$\pm$0.001 of the Type-1 AGN and the Type-2 AGN, respectively. Meanwhile, effects of 
different central AGN activities should be considered, because BH mass is one of fundamental 
parameters relative to AGN activities. Here, the narrow emission line flux rations of core 
component of [O~{\sc iii}]$\lambda5007$\AA~ to narrow H$\beta$ (O3HB) and of 
[N~{\sc ii}]$\lambda6583$\AA~ to narrow H$\alpha$ (N2HA) are mainly considered to trace central 
AGN activities, the commonly applied ratios in the well-known Baldwin-Phillips-Terlevich 
diagrams \citep{bpt, kb01, ka03a, kb06, kb19, zh20}. Different distributions of O3HB and 
N2HA are shown in left middle two panels of Fig.~\ref{dis} with mean values of $\log(O3HB)$ 
($\log(N2HA)$) about 0.754$\pm$0.008 and 0.454$\pm$0.003 (-0.091$\pm$0.004 and -0.057$\pm$0.002) 
of the Type-1 AGN and the Type-2 AGN, respectively. Moreover, considering emission luminosity 
also sensitively depending on central AGN activities, effects of different [O~{\sc iii}] line 
luminosity $L_{O3}$ of core component are also well checked. Different distributions of 
$L_{O3}$ are shown in the bottom left panel of Fig.~\ref{dis}, with mean values of 
$\log(L_{O3}/({\rm erg/s}))$ about 40.945$\pm$0.008 and 40.282$\pm$0.007 of the Type-1 AGN 
and the Type-2 AGN, respectively. Uncertainty of each mean value of each distribution in 
the left panels of Fig.~\ref{dis} is simply estimated by the bootstrap method with 1000 loops.

	Before proceeding further, besides the distributions of O3HB and N2HA in left panels of 
Fig.~\ref{dis}, it is interesting to show properties of the collected Type-1 AGN and Type-2 AGN 
in the well-known BPT diagram in left panel of Fig.~\ref{bpt}, to show different locations 
between Type-1 AGN and Type-2 AGN in the BPT diagram. There are quite different central positions 
(marked as thick pluses) for the Type-1 AGN and the Type-2 AGN in the BPT diagram, to provide 
further clues to discuss effects of different central AGN activities on our final results on 
the different stellar velocity dispersions between Type-1 AGN and Type-2 AGN. Meanwhile, through 
the strip marked by dashed cyan lines with width about 0.1\ in the BPT diagram in left panel 
of Fig.~\ref{bpt}, the 1044 Type-1 AGN and the 4202 Type-2 AGN lying in the strip are collected 
to simply check effects of AGN activities on stellar velocity dispersions comparisons. The stellar 
velocity dispersion comparisons are shown in right panel of Fig.~\ref{bpt} of the 1044 Type-1 AGN 
and the 4202 Type-2 AGN in the strip, with mean $\log(\sigma)$ about 
$\log(\sigma/{\rm km\cdot~s^{-1}})\sim2.196\pm0.005$ ($\sigma~\sim~157\pm2 {\rm km\cdot~s^{-1}}$) 
and $\log(\sigma/{\rm km\cdot~s^{-1}})\sim2.101\pm0.002$ ($\sigma~\sim~126\pm1 {\rm km\cdot~s^{-1}}$) 
for the 1044 Type-1 AGN and the 4202 Type-2 AGN in the strip, respectively. The uncertainties of 
the mean values are determined by the bootstrap method with 1000 loops. Considering the measured 
stellar velocity dispersions in Type-1 AGN 4\% higher than the intrinsic values, the corrected 
mean stellar velocity dispersion is about $\sigma~\sim~151\pm2 {\rm km\cdot~s^{-1}}$ in the 1044 
Type-1 AGN, still indicating higher stellar velocity dispersions in Type-1 AGN. Different 
strips in the BPT diagram can lead to the similar higher stellar velocity dispersions in Type-1 
AGN covered in the strips. Here, we do not show more results in different strips, however, the 
results indicate more detailed discussions are necessary on effects of AGN activities.

	In order to ignore the necessary effects of different distributions of redshift ($z$), 
narrow line flux ratios (O3HB and N2HA) and line luminosity of core components of [O~{\sc iii}] 
line ($L_{O3}$) on direct $\sigma$ comparisons, the most convenient way is to create two subsamples 
of Type-1 AGN and Type-2 AGN having the same distributions of $z$, O3HB, N2HA and $L_{O3}$. 
Based on the measured $z$, O3HB, N2HA and $L_{O3}$ of the 6260 Type-1 AGN and the 15353 Type-2 
AGN in the main samples, it is easy to create a subsample of Type-2 AGN having the same 
distributions of $z$, O3HB, N2HA and $L_{O3}$ as those of the Type-1 AGN in the subsample, 
through finding minimum parameter distance $D_p~<~D_{cri}$ calculated as 
\begin{equation}
\begin{split}
D_{p,~i}~&=~D_{z,~i}~+~D_{O3HB,~i}~+~D_{N2HA,~i}~+~D_{L_{O3},~i} \\
    &=~(\frac{z_{1, i} - z_2}{sca_z})^2~+~(\frac{O3HB_{1,i} - O3HB_{2}}{sca_{O3HB}})^2\\
    &\ \ ~+~(\frac{N2HA_{1,i} - N2HA_{2}}{sca_{N2HA}})^2\\
    &\ \ ~+~(\frac{L_{O3,~1,~i} - L_{O3,~2}}{sca_{L_{O3}}})^2\ \ \ \ \ for\ i=1,\dots, N_1
\end{split}
\end{equation}
where $z_{1,~i}$, $O3HB_{1,~i}$, $N2HA_{1,~i}$ and $L_{O3,~1,~i}$ mean parameters of the $i$th 
Type-1 AGN in the main sample ($N_1~=~6260$), $z_2$, $O3HB_{2}$, $N2HA_{2}$ and $L_{O3,~2}$ 
mean parameters of all $N_2~=~15353$ ($N_2~>~N_1$) Type-2 AGN in the main sample, $sca_z$, 
$sca_{O3HB}$, $sca_{N2HA}$ and $sca_{L_{O3}}$ are scale factors leading to $D_z$, $D_{O3HB}$, 
$D_{N2HA}$ and $D_{L_{O3}}$ not much different in quantity, and $D_{cri}$ means a critical 
value to prevent high $D_p$ leading to much different distributions of $z$, O3HB, N2HA and 
$L_{O3}$ between the created final two subsamples. Then, based on $sca_z~\sim~0.0015$ and 
$sca_{O3HB}~\sim~0.006$, $sca_{N2HA}~\sim~0.0035$ and $sca_{L_{O3}}~\sim~0.02$ and 
$D_{cri}~\sim~80$, one subsample of 1445 Type-1 AGN and one subsample of 1445 Type-2 AGN are 
created, which have the same distributions of $z$, O3HB, N2HA and $L_{O3}$ with significance 
levels higher than 99.5\% through the two-sided Kolmogorov-Smirnov statistic technique. 
Certainly, each object in the main samples is selected once in the two subsamples. The 
distributions of $z$, O3HB, N2HA and $L_{O3}$ for the AGN in the subsamples are shown in 
right panels of Fig.~\ref{dis}. The parameters, such as $\sigma$, $z$, O3HB, N2HA and $L_{O3}$, 
etc., of the 1445 Type-1 AGN and the 1445 Type-2 AGN in the subsamples are listed in Table~1 and 
in Table~2.

	Similar as Fig.~\ref{sigma12} but for the 1445 Type-1 AGN and the 1445 Type-2 AGN in the 
subsamples, stellar velocity dispersion comparisons are shown in Fig.~\ref{dis2}. After 
considering necessary effects, statistically larger $\sigma$ can also be clearly confirmed in 
Type-1 AGN with mean $\log(\sigma/{\rm km\cdot~s^{-1}})~\sim~2.166\pm0.005$ 
($\sigma~\sim~147\pm2{\rm km\cdot~s^{-1}}$) than in Type-2 AGN with mean 
$\log(\sigma/{\rm km\cdot~s^{-1}})~\sim~2.115\pm0.004$ ($\sigma~\sim~130\pm2{\rm km\cdot~s^{-1}}$). 
Considering the measured stellar velocity dispersions in Type-1 AGN 4\% higher than the intrinsic 
values, the corrected mean stellar velocity dispersions is about $\sigma~\sim~142\pm2 {\rm km\cdot~s^{-1}}$ 
in the 1445 Type-1 AGN in the subsample, still indicating $(9\pm3)$\% higher stellar velocity dispersions 
in Type-1 AGN. And the Students T-statistic technique is applied to determine the different mean 
values of $\log(\sigma)$ shown in Fig.~\ref{dis2} with confidence level higher than 10sigma between 
the Type-1 AGN and the Type-2 AGN.

	It is hard to reasonably explain the statistically larger $\sigma$ in Type-1 AGN than in 
Type-2 AGN, unless to assume that there were some lost Type-2 AGN with larger $\sigma$ and/or assume 
that there were some lost Type-1 AGN with smaller $\sigma$. Larger $\sigma$ commonly indicates stronger 
host galaxy contributions, indicating it is hard to miss Type-2 AGN with larger $\sigma$, due to their 
expected more apparent absorption features. Meanwhile, it could be expected to miss Type-1 AGN with 
smaller $\sigma$, due to weaker host galaxy contributions but stronger central AGN activities. The 
expected strong AGN activities can clearly lead the lost Type-1 AGN with smaller intrinsic $\sigma$ 
to be a true Type-1 AGN but without measured stellar velocity dispersions, quite similar as the Type-1 
AGN without measured $\sigma$ in the parent sample. As discussed results in subsection 4.3, the Type-1 
AGN without measured $\sigma$ could have statistically larger (at least not smaller) $\sigma$ than 
the Type-1 AGN with well measured $\sigma$. Therefore, statistically larger $\sigma$ is robust enough 
in Type-1 AGN than in Type-2 AGN.

\begin{figure}
\centering\includegraphics[width = 8cm,height=5cm]{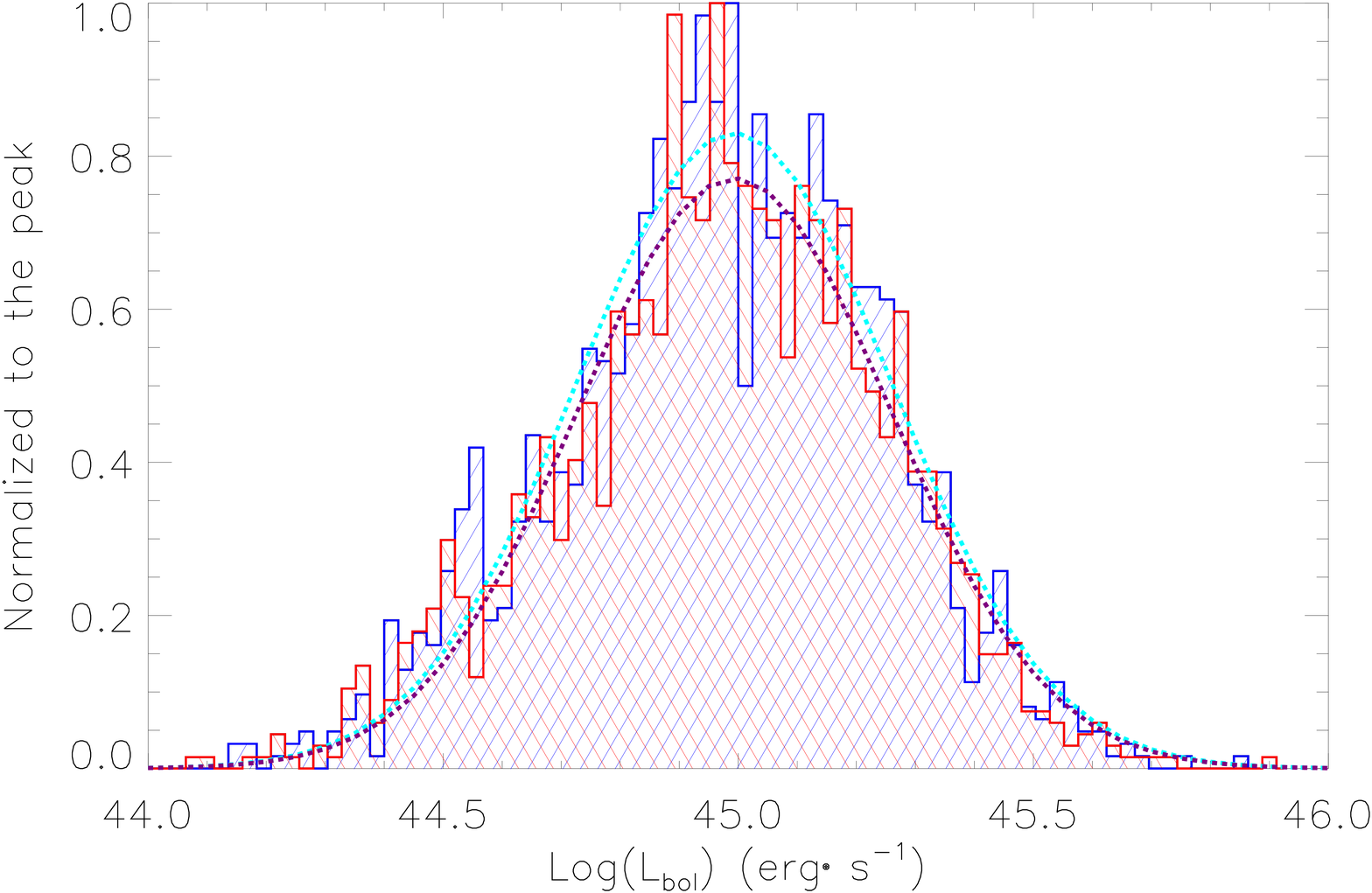}
\caption{Bolometric luminosity distributions of the 1445 Type-1 AGN (histogram filled by blue lines) 
and the 1445 Type-2 AGN (histogram filled by red lines) in the subsamples, which have the same 
distributions of $z$, O3HB, N2HA and $L_{O3}$. Dashed lines in purple and in cyan represent the 
corresponding best fitting Gaussian profiles for the distributions of the Type-2 AGN and of the 
Type-1 AGN, respectively.}
\label{Lbol}
\end{figure}

	Before the end of the subsection, three points are noted. First, effects of different 
aperture sizes are not considered on measured $\sigma$, because the same effects can be confirmed 
for the 1445 Type-1 AGN and the 1445 Type-2 AGN in the subsamples with the same redshift 
distributions, indicating few effects of different aperture sizes on the results shown in 
Fig.~\ref{dis2}. Second, effects of host galaxy morphologies (including contributions of bars 
and/or disks of host galaxies) are not considered on the measured $\sigma$, because barred 
galaxies, unbarred galaxies, merging galaxies, and those hosting pseudo-bulges do not represent 
outliers in \msig relations as discussed in more recent \citet{bt15, bt21} and as shown results 
in \citet{bb17}. Third, effects of inclination are not considered on the measured stellar 
velocity dispersions. As discussed in \citet{bh14} and then followed in \citet{sc19}, there 
are apparent effects of galaxy orientation on measured stellar velocity dispersions, due to 
contributions of disk rotating components. However, more recently, from a Hubble 
Space Telescope snapshot imaging survey, \citet{kh21} have shown\footnote{In 2018, \citet{bg18} 
have shown that host galaxies of the majority of Type 1 AGN are elliptical and/or compact 
galaxies, while host galaxies of Type 2 AGN present more scatters, but through samples of high 
redshift Type-1 AGN and Type-2 AGN with redshift larger than 0.3 and smaller than 1.1. 
Therefore, in the manuscript, results in \citet{kh21} for low redshift AGN are preferred and 
mainly considered.} that Type-1 and Type-2 AGN are almost indistinguishable in terms of their 
Hubble type distributions, through Swift-BAT unbiased X-ray selected AGN with redshift smaller 
than 0.1 and with bolometric luminosity around $10^{43-46}{\rm erg\cdot s^{-2}}$. Based on the 
reported luminosity $L_{o3}$ of core components of [O~{\sc iii}] lines of the AGN in the subsamples, 
the bolometric luminosities $L_{bol}\sim10\times L_{5100\textsc{\AA}}$ \citep{rg06, du20, nh20} 
can be estimated by the reported correlation between continuum luminosity $L_{5100\textsc{\AA}}$ 
and $L_{o3}$ in \citet{zh17}, and shown in Fig.~\ref{Lbol} for the Type-1 AGN and Type-2 AGN 
in the subsamples, which are well comparable to the bolometric luminosities of AGN in \citet{kh21}. 
Therefore, considering the results in \citet{kh21} that there is negligible difference in terms 
of Hubble type between Type-1 and Type-2 AGN, there are no statistical inclination effects on 
our final stellar velocity dispersion comparisons between Type-1 AGN and Type-2 AGN in the 
manuscript.

	Based on the results above, statistically larger $\sigma$ can be well confirmed in the 
Type-1 AGN than in the Type-2 AGN with significance level higher than 10sigma, after considerations 
of necessary effects. Therefore, unless there was stable evidence to support different \msig relations 
applied to determine central BH masses or to support quite different evolution histories in Type-1 
AGN and in Type-2 AGN, larger stellar velocity dispersions in the Type-1 AGN than in the Type-2 AGN 
leads to an interesting challenge to the Unified Model of AGN.

\section{Summaries and Conclusions}

   The main summaries and conclusions are as follows. 
\begin{itemize}   
\item So-far the largest sample of 6260 low redshift Type-1 AGN ($z~<~0.3$), about 85 times 
	larger than the more recent sample of Type-1 AGN with measured spatially resolved 
	stellar velocity dispersions $\sigma_{bt}$ in \citet{bt15, bt21}, have their stellar 
	velocity dispersions $\sigma$ measured through absorption features around 4000\AA. 
	And the measured $\sigma$ are well consistent with the $\sigma_{bt}$ for the Type-1 
	AGN with apparent absorption features around 4000\AA.
\item Meanwhile, almost all the low redshift Type-2 AGN ($z~<~0.3$) in SDSS DR12 have the 
	stellar velocity dispersions $\sigma$ measured through absorption features around 
	4000\AA. And the measured $\sigma$ are well consistent with the SDSS provided 
	dispersions of the Type-2 AGN.
\item Based on series of artificial spectra created by Type-2 AGN spectrum plus contributions 
	from both AGN continuum emissions and broad line emissions, contributions from both AGN 
	continuum emissions and broad line emissions can also lead to reliable measured stellar 
	velocity dispersions through the absorption features around 4000\AA, however, strong AGN 
	continuum emissions can lead to part of Type-1 AGN of which stellar velocity dispersions 
	cannot be measured, and lead to the measured stellar velocity dispersions 
	about 4\% higher than the intrinsic values.
\item Although half of low redshift Type-1 AGN, about 6082 Type-1 AGN, have not apparent 
	absorption features around 4000\AA~, leading the half of low redshift Type-1 AGN 
	without measured stellar velocity dispersions, properties of the mean spectra of the 
	6260 Type-1 AGN with measured $\sigma$ and the 6082 Type-1 AGN without measured 
	$\sigma$ can be applied to confirm that the 6082 Type-1 AGN without measured $\sigma$ 
	are mainly due to weak absorption features overwhelmed in the emission features of 
	central AGN activities.
\item Based on the positive correlation between $\sigma$ and [O~{\sc iii}] line luminosity, 
	the half of low redshift Type-1 AGN without apparent absorption features around 4000\AA~ 
	should have intrinsic stellar velocity dispersions statistically larger (at least not 
	smaller) than the 6260 Type-1 AGN with measured stellar velocity dispersions, due to 
	statistically higher [O~{\sc iii}] line luminosities of the half of low redshift Type-1 
	AGN without apparent absorption features around 4000\AA.
\item Based on the measured stellar velocity dispersions $\sigma$ of the largest sample of 
	Type-1 AGN and the largest sample of Type-2 AGN in SDSS DR12, direct $\sigma$ comparisons 
	can lead to statistically larger $\sigma$ in the Type-1 AGN than in the Type-2 AGN, 
	without considerations of necessary further effects.
\item Even after considering necessary effects of different evolution histories and central 
	AGN activities on direct $\sigma$ comparisons between Type-1 AGN and Type-2 AGN, 
	such as the effects of different redshift, different [O~{\sc iii}] line luminosities 
	of core components and different narrow line ratios of O3HB and N2HA, 
	statistically $(9\pm3)$\% larger $\sigma$ can be well confirmed in the Type-1 AGN 
	than in the Type-2 AGN with significance level higher than 10sigma.
\item Unless there was strong evidence to support different \msig relations or to support quite 
	different evolution histories between Type-1 AGN and Type-2 AGN, the statistically 
	larger $\sigma$ in Type-1 AGN provides an interesting but strong challenge to the 
	Unified model of AGN.
\end{itemize}

\begin{table*}
\caption{Parameters of the 1445 Type-1 AGN in the subsample}
\begin{center}
\begin{tabular}{cccccc|cccccc}
\hline\hline
mpf & $z$ & $\sigma$ & $L_{O3}$ & O3HB & N2HA & mpf & $z$ & $\sigma$ & $L_{O3}$ & O3HB & N2HA \\
\hline
0266-51602-0239   &   0.063   &   106$\pm$11   &   40.22   &   0.705   &   -0.33   &   
0267-51608-0300   &   0.067   &   99$\pm$13   &   40.32   &   0.583   &   -0.03   \\
0270-51909-0266   &   0.110   &   124$\pm$13   &   41.06   &   1.045   &   -0.11   &   
0270-51909-0429   &   0.128   &   83$\pm$14   &   39.87   &   0.104   &   -0.29   \\
0271-51883-0200   &   0.061   &   140$\pm$9   &   40.57   &   0.929   &   0.162   &   
0271-51883-0322   &   0.086   &   140$\pm$9   &   41.10   &   1.032   &   0.181   \\
0272-51941-0329   &   0.078   &   135$\pm$10   &   41.21   &   0.936   &   -0.01   &   
0273-51957-0460   &   0.096   &   131$\pm$15   &   41.28   &   1.013   &   0.005   \\
0273-51957-0579   &   0.131   &   238$\pm$22   &   41.91   &   1.072   &   0.022   &   
0274-51913-0141   &   0.138   &   111$\pm$10   &   41.28   &   0.988   &   0.098   \\
\hline
\end{tabular}
\end{center}
\tablecomments{
The first column and the seventh column show the information of SDSS PLATE-MJD-FIBERID. 
The second column and the eighth column show the redshift. The third column and the ninth column 
show the measured $\sigma$ in the unit of ${\rm km\cdot~s^{-1}}$ through the absorption features 
around 4000\AA. The fourth column and the tenth column show the $\log(L_{O3})$ of the core 
components of [O~{\sc iii}]$\lambda5007$\AA~ in the unit of ${\rm erg\cdot~s^{-1}}$. The fifth 
column and the eleventh column show the $\log(O3HB)$. The sixth and twelfth columns show the 
$\log(N2HA)$.\\
\ \ \ \\
Table 1 is published in its entirety in the machine-readable format.
}
\end{table*}

\begin{table*}
\caption{Parameters of the 1445 Type-2 AGN in the subsample}
\begin{center}
\begin{tabular}{cccccc|cccccc}
\hline\hline
mpf & $z$ & $\sigma$ & $L_{O3}$ & O3HB & N2HA & mpf & $z$ & $\sigma$ & $L_{O3}$ & O3HB & N2HA \\
\hline
0266-51630-0147   &   0.030   &   99$\pm$7   &   39.92   &   1.007   &   -0.05   &   
0266-51630-0392   &   0.122   &   124$\pm$18   &   41.21   &   0.571   &   -0.04   \\
0269-51910-0105   &   0.184   &   126$\pm$21   &   40.75   &   0.531   &   -0.04   &   
0269-51910-0168   &   0.176   &   131$\pm$21   &   40.78   &   0.889   &   -0.03   \\
0271-51883-0178   &   0.181   &   124$\pm$15   &   41.76   &   0.814   &   -0.19   &   
0272-51941-0529   &   0.177   &   208$\pm$24   &   40.91   &   0.871   &   -0.04   \\
0274-51913-0115   &   0.095   &   107$\pm$10   &   40.38   &   0.633   &   -0.24   &   
0274-51913-0230   &   0.077   &   115$\pm$7   &   41.19   &   0.650   &   -0.24   \\
0275-51910-0438   &   0.150   &   92$\pm$12   &   41.27   &   0.681   &   -0.15   &   
0276-51909-0147   &   0.108   &   117$\pm$18   &   40.12   &   0.443   &   -0.29   \\
\hline
\end{tabular}
\end{center}
\tablecomments{The first column and the seventh column show the information of SDSS PLATE-MJD-FIBERID. 
The second column and the eighth column show the redshift. The third column and the ninth column 
show the measured $\sigma$ in the unit of ${\rm km\cdot~s^{-1}}$ through the absorption features 
around 4000\AA. The fourth column and the tenth column show the $\log(L_{O3})$ of the core components 
of [O~{\sc iii}]$\lambda5007$\AA~ in the unit of ${\rm erg\cdot~s^{-1}}$. The fifth column and the 
eleventh column show the $\log(O3HB)$. The sixth and twelfth columns show the $\log(N2HA)$. \\ 
\ \ \ \\
Table 2 is published in its entirety in the machine-readable format.
}
\end{table*}

\section*{Acknowledgements}
Zhang gratefully acknowledge the anonymous referee for giving us constructive 
comments and suggestions to greatly improve the paper. Zhang gratefully 
thanks the kind financial support from Nanjing Normal University and the 
kind grant support from NSFC-12173020. This manuscript has made use of the 
data from the SDSS projects. The SDSS-III web site is http://www.sdss3.org/. SDSS-III 
is managed by the Astrophysical Research Consortium for the Participating 
Institutions of the SDSS-III Collaborations. The manuscript has made use 
of the data from the Indo-U.S. Coude Feed Spectral Library 
(\url{https://www.noao.edu/cflib/}) which consists of spectra for 1273 
stars obtained with the 0.9m Coude Feed telescope at Kitt Peak National Observatory.


\end{document}